\documentclass[sn-mathphys-ay]{sn-jnl}

\usepackage{fullpage, amsfonts, euscript, times, graphicx, hyperref, float}

\usepackage{physics}

\graphicspath{{./include/}}

\usepackage{xcolor}
\newcommand{\bit}{\begin{itemize}}
\newcommand{\eit}{\end{itemize}}
\newcommand{\beq}{\begin{equation}}
\newcommand{\eeq}{\end{equation}}
\newcommand{\Eref}{Eqn.~\eqref}
\newcommand{\Fref}{Fig.~\ref}

\usepackage{xspace}

\newcommand{\lb}{\left(}
\newcommand{\rb}{\right)}

\renewcommand{\vec}{\vb}

\usepackage{graphicx}%
\usepackage{multirow}%
\usepackage{amsmath,amssymb,amsfonts}%
\usepackage{amsthm}%
\usepackage{mathrsfs}%
\usepackage[title]{appendix}%
\usepackage{xcolor}%
\usepackage{textcomp}%
\usepackage{manyfoot}%
\usepackage{booktabs}%
\usepackage{algorithm}%
\usepackage{algorithmicx}%
\usepackage{algpseudocode}%
\usepackage{listings}%

\theoremstyle{thmstyleone}%
\theoremstyle{thmstyletwo}%

\theoremstyle{thmstylethree}%

\begin{document}

\title[How cells stay together]{
    How cells stay together;
    a mechanism for maintenance of a robust cluster explored by local and nonlocal continuum models}


\author*[1]{\fnm{Andreas} \sur{Buttensch\"{o}n}}\email{andreas.buttenschoen@umass.edu}

\author[2]{\fnm{Shona} \sur{Sinclair}}\email{shona.sinclair@shaw.ca}

\author[2]{\fnm{Leah} \sur{Edelstein-Keshet}}\email{keshet@math.ubc.ca}

\affil*[1]{\orgdiv{Department of Mathematics and Statistics}, \orgname{University of Massachusetts}, \orgaddress{\street{710 N.\ Pleasant St}, \city{Amherst}, \postcode{01003}, \state{MA}, \country{United Stated of America}}}

\affil[2]{\orgdiv{Department of Mathematics}, \orgname{University of British Columbia}, \orgaddress{\street{1984 Mathematics Road}, \city{Vancouver}, \postcode{V6T 1Z2}, \state{BC}, \country{Canada}}}


\abstract{
 Formation of organs and specialized tissues in embryonic development requires
 migration of cells to specific targets. In some instances, such cells migrate
 as a robust cluster. We here explore a recent local approximation of nonlocal
 continuum models by Falc\'o, Baker, and Carrillo (2023). We apply their
 theoretical results by specifying biologically-based cell-cell interactions,
 showing how such cell communication results in an effective
 attraction-repulsion Morse potential. We then explore the clustering
 instability, the existence and size of the cluster, and its stability. We also
 extend their work by investigating the accuracy of the local approximation
 relative to the full nonlocal model.
}

\keywords{Stable cell cluster, attractant-repellent chemotaxis, Morse potential, nonlocal PDE, local PDE approximation}

\pacs[MSC Classification]{95B05}

\maketitle

\bmhead{Acknowledgements}

LEK is funded by a Natural Science and Engineering Research Council (NSERC,
Canada) Discovery Grant. SRS was funded by an  NSERC undergraduate summer
research assistantship (USRA) in  2023. We are grateful to Paul Kulesa and
Jennifer C. Kasemeier-Kulesa (Notre Dame University and Stowers Institute) for
discussions and for the original biological motivation that led to the topic of
robust cell clusters. We also  wish to acknowledge the (1995) UBC PhD thesis of
Alex Mogilner where links between chemotaxis, Morse potentials, and regimes of
behaviour were originally shown.

\section{Introduction}

Migration of clusters of tumor cells found by \cite{friedl1995migration}  introduced the importance of collective cell behaviour and migration. However, cell clusters form (and migrate collectively) in many normal developmental and morphogenetic processes \citep{weijer2009collective}. This is true of the post-aggregation ``slug'' stage of social amoebae such as Dichtyostelium discoideum (DD) \citep{bonner1998way} and the primary lateral line primordial (PLLP), a cluster of about 100 cells that deposits sensory organs along the length of a zebrafish embryo \citep{chitnis2012building}.
Long range migration by a cluster of cells is essential in the formation of the sympathetic nervous system, as studied in \cite{Kasemeier2015} for chick embryos. Cell clusters can depend on actual contact, adhesion, and short or long-lived junctions formed between cells \citep{friedl2017tuning}. But in the case of the primary sympathetic ganglion (PSG) cluster described in \cite{Kasemeier2015}, the cells form a ``swarm'' and avoid adhering to one another.

A question arising in these experiments is what determines the observed cell cluster sizes, and in particular how cluster size is  related to cell-cell interactions.
From spatial RNAseq experiments in the PSG, it is known that the cells in the migrating cluster interact via chemical signalling \citep{Kasemeier2015}, but whether the chemicals are attractants, repellents, or both, is unclear. More generally, an interesting biological problem is to determine what combination of such chemicals (with various rates of diffusion, secretion, and decay) would be consistent with the formation and maintenance of robust cell clusters. This general question motivates our paper.

 Models of collective cell behaviour have utilized the cellular Potts Model \citep{maree1999migration} and a spheroidal (3D) cell-based computation of the DD slug \citep{palsson2000model}, and of the PLLP \citep{knutsdottir2017polarization}, and other custom-built 2D cell migration models for neural crest cells \citep{merchant2018rho}. See review in \cite{buttenschon2020bridging}.
Comparison of continuum and agent-based models in cell populations are found in \cite{byrne2009individual,chaplain2020bridging}.

 Non-local equations have frequently been used to describe cell densities. These include models that have non-local interactions a-priori such as the cell-cell adhesion model of \cite{armstrong2006continuum}, the non-local chemotaxis model by \cite{othmer2002diffusion}, or   models in which the non-local term appears due to a ``model reduction'' technique (e.g., quasi steady-state assumptions, as in \cite{knutsdottir2017polarization,mogilner1995PhDThesis}).
 A popular reduction of non-local models to local models has been considered in the literature by  \cite{murray2003mathematical}, by \cite{gerisch2008mathematical}, and recently by \cite{Falco2023}. In this reduction, the limits of integration are formally taken to zero, so that fully local equations are found. These local equations can give powerful insights and connections to well studied local equations
 e.g.\ Cahn-Hilliard or Allen-Cahn
 \citep{bernoff2016biological}.


While much of our paper rests on the results of \cite{Falco2023}, we here pose several new questions of both mathematical and biological flavour:

\begin{enumerate}

\item What biological processes give rise to cell-cell interactions? (Rephrased, where do the kernels describing cell ``potentials''  come from?)


\item Under what conditions (on the kernels or the parameters) would a single cell species create a stable cluster that is maintained?

\item How does the cluster size (radius) depend on the cell interaction parameters?

\item For a given underlying cell signaling model, can we create a ``map'' in parameter space that shows the possible regimes of behaviour of a population of such cells? What behaviours can occur other than clustering?

\item How does a local (PDE) approximation compare with the full nonlocal (integro-PDE) model?

\item How do the predictions of the continuum models compare with agent-based models with similar mechanisms?



\end{enumerate}

More broadly, we espouse the idea of using several kinds of model simplifications to gain initial understanding of processes, before attempting more complex scenarios and adding details. Indeed, we make several simplifications to obtain insight on the migration of the PSG cluster: (1) We first drop the analysis of migration and consider the problem of cluster existence, shape, and stability. (2) We carry out analysis in a 1D geometry, where results are simplest and most intuitive. (3)  We assume one species of identical cells.

In what follows, we will examine both nonlocal models and their local approximation, applying previous results of \cite{Falco2023,mogilner1995PhDThesis} to the questions posed above. 

\section{The nonlocal model and its approximation by a local PDE}

To keep this paper self-contained, we first provide a pedagogical summary of part of \cite{Falco2023}, listing some basic results from that paper.

Let $\rho(x,t)$ denote density of cells, at position $x$ and time $t$. Ignoring random cell motion, the full nonlocal model in generality is
\beq\label{eq:Carrillo1clusterIPDE}
    \frac{\partial \rho}{\partial t}= -\nabla \cdot( \rho \vec{v}), \quad \mbox{where} \quad \vec{v}=\vec{F},\quad \vec{F}= - \nabla W,\quad W= (K*\rho),
\eeq
where $W$ represents a potential induced
by a cell  on its neighboring cells at some position $\vec{x}$ at
time $t$. The velocity $\vec{v}$ is proportional to a force $\vec{F}$ in the low Reynold's number regime of cells; here the drag coefficient has been scaled to 1. The force is assumed to be the gradient of a potential induced by the cells. $K*\rho$ is a convolution that sums up all distance-dependent cellular interactions.

Here we consider the spatial 1D case, where the model reduces to
%
\beq\label{eq:Carrillo1clusterIPDE1D1}
    \frac{\partial \rho}{\partial t}= -\frac{\partial}{\partial x} ( \rho v), \quad v= - \frac{\partial (K*\rho)}{\partial x},
\eeq
where
\[
W := K*\rho = \int_{-\infty}^{\infty} K(x-s) \rho (x,t) dx=\int_{-\infty}^{\infty} K(z) \rho (x-z,t) dz
\]
is a convolution over the whole real line, $-\infty<x<\infty $. 
We will take the simplest case, where $K$ is a symmetric kernel, i.e.\
cells interact in a symmetric way, making $K$ an even function of space,
$K(x) = K(-x)$, so cells induce a potential that is symmetric about
their position.
This general problem was one of the topics studied in \cite{Falco2023}. We wrote the convolution in two forms, as the latter is useful to our next step.

As in \cite{Falco2023}, we consider a local approximation for the nonlocal 1D model \eqref{eq:Carrillo1clusterIPDE1D1}.   To do so, we transformed the integration variable ($z=x-s$) as shown above.
Expand the term $\rho (x-z,t)$ in a Taylor series about $x$ to obtain a sum of  partial derivatives $\rho, \rho_x, \rho_{xx}$ etc. The coefficients of such terms are proportional to moments of the kernel, that is
\begin{equation}
\label{eq: Coef a0,a2}
a_0=\lim_{u\to\infty}2 \int_0^u  K(z) dz,\quad
a_2=\lim_{u\to\infty}\frac12\cdot2 \int_0^u z^2 K(z) dz.
\end{equation}
The even symmetry of the potential implies that $a_1=0$. We keep terms up to order $z^2$. Then the local approximation is
\beq
\label{eq:Carrillo1clusterAproxPDE1D}
\frac{\partial \rho}{\partial t}= -\frac{\partial}{\partial x} ( \rho v), \quad \mbox{where} \quad v= - \frac{\partial }{\partial x} (a_0 \rho + a_2 \rho_{xx}).
\eeq

\subsection{Results from the Falco et al model}

We now mention several results from \cite{Falco2023}.
\begin{enumerate}
    \item The problem \eqref{eq:Carrillo1clusterIPDE1D1} with Neumann BCs can be solved using a variational approach, similar to a Cahn-Hilliard equation \citep{Falco2023,elliott1996cahn}, with a ``Free energy'',

\[
\Phi=\frac12 \int_{-\infty}^\infty a_0 \left|\frac{\partial\rho}{\partial x}\right|^2 - a_2 \rho^2 dx.
\]
    \item Solving for steady states of \eqref{eq:Carrillo1clusterIPDE1D1} is equivalent to finding critical points of the free energy $\Phi$.
    \item For a compact ``cluster'' solution with density $\rho \ge 0$ on $-b\le x\le b$, the minimum free energy is obtained when a ``zero contact angle'' is satisfied at the ``edges'' of the cluster, that is when
    \begin{equation}\label{eq:FalcaoBCs}
    \rho(x)=0, \quad \rho'(x)=0, \quad \text{at} \quad x=\pm b
        \end{equation}
    where $b$ is the ``radius'' of the cluster.
\end{enumerate}
Hence, in order to find a (local) approximation for the stable steady state cluster solutions, it suffices to solve the (local) PDE \eqref{eq:Carrillo1clusterAproxPDE1D} with the boundary conditions \eqref{eq:FalcaoBCs} and the above parameters $a_0, a_2$.

\subsection{Steady state of the local problem}

The steady state(s) of \eqref{eq:Carrillo1clusterAproxPDE1D} satisfy
\beq
\label{eq:SSclusterAproxPDE1D_0}
0= \frac{\partial}{\partial x} \left( \rho \frac{\partial }{\partial x} (a_0 \rho + a_2 \rho_{xx})\right).
\eeq
\subsubsection{Homogeneous steady state}

From \eqref{eq:SSclusterAproxPDE1D_0}, it is clear that there is always a homogeneous steady state (HSS) solution on a closed interval, namely
$\rho_0=M/L$ where $L$ is the size of the interval and $M$ is the total mass of the cells. The stability of this HSS is easily deduced with linear stability analysis, see below.

\subsubsection{Single cluster steady state}
We seek the more interesting case of a spatially nonhomogeneous steady state solution.
To do so, integrate \eqref{eq:SSclusterAproxPDE1D_0}  once. Then using the boundary conditions \eqref{eq:FalcaoBCs} results in
\beq
\label{eq:SSclusterAproxPDE1D_1}
0= \left( \rho \frac{\partial }{\partial x} (a_0 \rho + a_2 \rho_{xx})\right).
\eeq
We seek a single-cluster solution with compact support $\rho\ne 0$ on $-b <x<b$, so for this interval, we can divide both sides by $\rho$ and integrated once more, to obtain the PDE for the cluster density,
\beq
\label{eq:SSclusterAproxPDE1D_2}
  a_0 \rho + a_2 \rho_{xx} = C,
\eeq
where $C$ is a constant to be determined. The PDE \eqref{eq:SSclusterAproxPDE1D_2} is now a simple second order linear ODE, whose solutions can be represented in the form of sine and cosine functions. By symmetry and by the boundary conditions, we can restrict attention to solutions of the form
\[
\rho(x)=\alpha \cos(\mu x),
\]
where $\alpha, \mu$ are parameters to be determined. Furthermore, for a single compact cluster, we seek one period of the above periodic function, so we restrict attention to solutions that are nonzero on a closed interval $|x|<b$, where $b>0$ is identified with the cluster radius (and hence $b=\pi/\mu$).

\begin{figure}[h]
    \centering
    \includegraphics[scale=0.8]{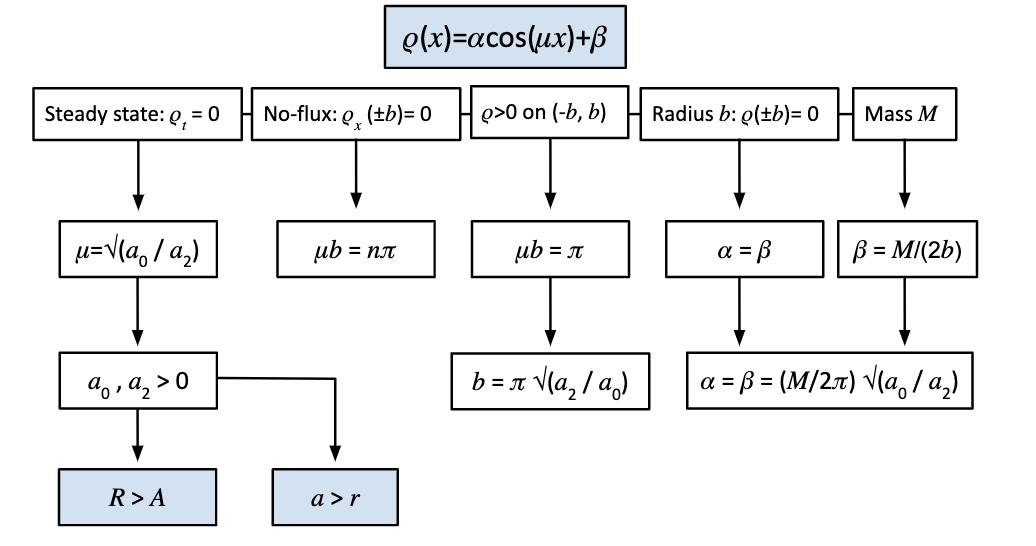}
    \caption{Flow chart depicting how each parameter constraint was deduced for the existence of a cluster in the 1D local PDE approximation for the cell density.
    }
    \label{fig:1-species-flowchart}
\end{figure}

In the flow chart of Fig.~\ref{fig:1-species-flowchart}, we provide the logic and conditions for existence of a steady state single-cluster solution. The steps are general, with the exception of the last (blue shaded) inequalities that will be discussed later on in the context of the Morse potential.
So far, our results are merely a summary of the general theory of \cite{Falco2023} for 1 species in 1D. The application to the biological example will result in several conditions on how cell signaling should be tuned to arrive at such a cluster solution.

\section{Stability}

\subsection{Stability of the homogeneous steady state}
We can state several general results about the stability of a spatially uniform steady state $\rho_0$ based on linear stability analysis (LSA).

\subsubsection{LSA for the nonlocal problem}
For the full nonlocal PDE \eqref{eq:Carrillo1clusterIPDE1D1}, the linearized version, obtained by setting $\rho(x,t)=\rho_0+\rho'(x,t)$ is
\[
  \frac{\partial \rho'}{\partial t} =   \rho_0 \frac{\partial^2}{\partial x^2}\lb  K*\rho'\rb ,
\]
where $\rho'(x,t)$ is a small spatially varying perturbation. Let  $\rho'(x,t)=\tilde{\rho}\exp(\lambda t) \exp(i q x)$ where $\lambda$ is the perturbation growth rate, $q$ is the perturbation wavenumber, and $\tilde{\rho}$ is a small amplitude. We seek conditions such that, for some wavenumber(s), the perturbations grow (Re$(\lambda)>0 )$. Substituting the form of the perturbations into the linearized PDE leads to the algebraic equation
\beq
\label{eq:NonlocalHSSStabil}
\lambda(q)=\rho_0 (-q^2) \hat{K}(q).
\eeq
where $\hat{K}(q)$ is the Fourier transform of the kernel $K(x)$. We can hence determine which wavenumbers would grow from the Fourier transform of the kernel of interest.

\subsubsection{LSA for the local problem}
Making the same assumptions about the perturbation $\rho'(x,t)$ and substituting into the linearized local PDE leads to the eigenvalue expression
\beq
\label{eq:localPDEeigenv}
\lambda= \rho_0 q^2
 \left( a_0  - a_2 q^2\right).
\eeq
The HSS can be destabilized to growing perturbations whenever the RHS is positive, i.e. when $0<q<\sqrt{a_0/a_2}.$
This mandates that either one of the two conditions below be satisfied:
\begin{equation}
 \label{eq:LSAcondsOn_a's}
a_0, a_2 >0 \quad \mbox{or} \quad a_0, a_2 <0
\end{equation}
Again, once a potential kernel is selected, these inequalities directly lead conditions on its moments, and hence on parameters that describe that potential.

\subsection{Stability of the nonhomogeneous (clustered) steady state}
This problem is more challenging, as it requires perturbation of the cluster solution. Here we will not carry out a full analysis, but rather determine stability  of the cluster using numerical simulations.

\section{Morse kernels}
We apply the above general theory of \cite{Falco2023} to a biologically relevant setting where cell signaling is based on known or plausible mechanism(s). We here introduce the Morse potentials, and then show that they capture known cell-cell signaling mechanism(s).

The Morse potential we consider combines repulsion and attraction, written as
\begin{equation}\label{eq:MorsePotential1}
 K(x)=R r\exp\lb -\left|\frac{x}{r}\right|\rb- A a \exp\lb -\left|\frac{x}{a}\right|\rb
\end{equation}
where $A,R \ge 0$ are (attraction, repulsion) force magnitudes and $a,r\ge 0$ are typical spatial scales of attraction and repulsion. This is an even function. It is to be interpreted as a symmetric ``potential'' induced by a single cell located at the origin on its neighbors at another location $x$. We have used the above notation to be consistent with \cite{mogilner1995PhDThesis} where forces (rather than potentials) were employed in a similar setting.
It will be convenient to define the following dimensionless ratios:
\[
C=\frac{R}{A},\quad \ell = \frac{a}{r}.
\]
Our results will be summarized as regimes in the $C\ell$ plane, where inequalities determine certain regions whose boundaries are expressed in terms of $C$ and $\ell$.

\subsection{Results for Morse potentials}

In 1D, the Morse potential \eqref{eq:MorsePotential1} generates a force-field of the form
\[
F_\text{Morse}= -\dd K(x)/ \dd x = \text{sign}(x) \left(R \exp\lb -\left|\frac{x}{r}\right|\rb - A  \exp\lb -\left|\frac{x}{a}\right|\rb \right).
\]
This is an odd function.
Intuitively, for cells to form a cluster, we require strong local repulsion (to avoid collapse) and more long-ranged attraction (to keep cells from fleeing to infinity). These conclusions will emerge from our analysis and are known from previous work for agent-based models (ABMs) by \cite{mogilner2003mutual, dOrsogna2006self}. This leads to the first two conditions,
\[
R>A,\ a>r \quad \Rightarrow \quad C>1,\ \ell>1.
\]
We can now apply various general results to the Morse potentials.

\subsubsection{Existence of a cluster}
We easily find from \eqref{eq: Coef a0,a2} that in the approximate local 1D PDE problem~\eqref{eq:Carrillo1clusterAproxPDE1D}, the coefficients of interest are
\begin{equation}
\label{eq: MorseLocalCoeffs}
        a_0=2(Rr^2-Aa^2), \quad   a_2=2(Rr^4-Aa^4).
\end{equation}

The inequalities $a_0, a_2>0$ imply that
\begin{equation}
\label{eq:TheTwoIneq}
Rr^2-Aa^2>0, \quad Rr^4-Aa^4>0, \quad
        \Rightarrow \quad C<\ell^2,\quad C<\ell^4.
\end{equation}
If both coefficients are negative, the inequalities are reversed. We later show that a \textbf{stable} cluster can only exist for the inequalities given in \eqref{eq:TheTwoIneq}, and not for the second case.

Collecting conditions so far, we have
\[
C>1, \quad \ell>1, \quad C<\ell^2,\quad C<\ell^4.
\]
These form the boundary curves of interest in the $C\ell$ parameter plane, shown in Figs.~\ref{Fig:ClusterSizes} and~\ref{Fig:KymographsSims}. We refer to behavioural regimes A-H as labeled in Fig.~\ref{Fig:ClusterSizes} and Fig.~\ref{Fig:KymographsSims}. The same set of curves were found in \cite{mogilner1995PhDThesis} (p.~103, Section~5.179 and Fig.~22) in calculation of a swarming instability.

\subsubsection{Cell density in the cluster}

For the Morse potentials, the density of cells in the cluster is
\beq
\label{eq: rho(x)}
    \rho(x)=\frac M{2\pi}\sqrt{\frac{Rr^2-Aa^2}{Rr^4-Aa^4}}\left(\cos\left(x\sqrt{\frac{Rr^2-Aa^2}{Rr^4-Aa^4}}\right)+1\right),
\eeq
where $M$ is the total mass of the cluster, obtained by integrating $\rho(x)$ over $[-b, b]$.  \Eref{eq: rho(x)} can be rewritten as
\beq
\label{eq: rho-C-ell}
    \rho(x)=\frac{M}{2\pi r}\sqrt{\frac{C-\ell^2}{C-\ell^4}}\left(\cos\left(\frac x r\sqrt{\frac{C-\ell^2}{C-\ell^4}}\right)+1\right).
\eeq
(Note that should we desire to scale space by the repulsive length scale $r$ and density by $M/r$, then only dimensionless quantities would remain in \eqref{eq: rho-C-ell}.)
We later refer to the spatial frequency $\omega$ and the dimensionless spatial frequency $\tilde{\omega}$ in \eqref{eq: rho(x)} and\eqref{eq: rho-C-ell} meaning
\begin{equation}
\label{eq:Frequencies}
\omega = \sqrt{\frac{Rr^2-Aa^2}{Rr^4-Aa^4}}, \quad
 \tilde{\omega} =  \sqrt{\frac{C-\ell^2}{C-\ell^4}}.
\end{equation}
Clearly, both $\omega$ and $\tilde{\omega}$ should be real valued for a cluster to exist.

\subsubsection{Cluster size}
The radius of the cluster, obtained by ensuring a one-period cosine solution is given by
\begin{equation}\label{Eqn:ClusterRadiusF}
b=\pi \sqrt{\frac{a_2}{a_0}}= \frac{\pi}{\omega} 
\quad \Rightarrow
 b=\, r \, \mathcal{F}, \quad \text{where  }  \mathcal{F} = \frac{\pi}{\tilde{\omega}} 
\end{equation}
for $\omega$ and $\tilde{\omega}$ in \eqref{eq:Frequencies}.
In \eqref{Eqn:ClusterRadiusF}, we defined a dimensionless quantity, $\mathcal{F}$.
Observe that $r$ and $b$ each carry units of length, and $r$ is a typical repulsive length scale.  Hence,  $\mathcal{F}$ is a ``dimensionless cluster size'', a constant that specifies the size of a cluster as a fold-multiple of the spatial scale $r$.
For a real-valued cluster solution to exist, $\mathcal{F}$ should be real.

%

\begin{figure}[!ht]\centering
    \includegraphics[width=\textwidth]{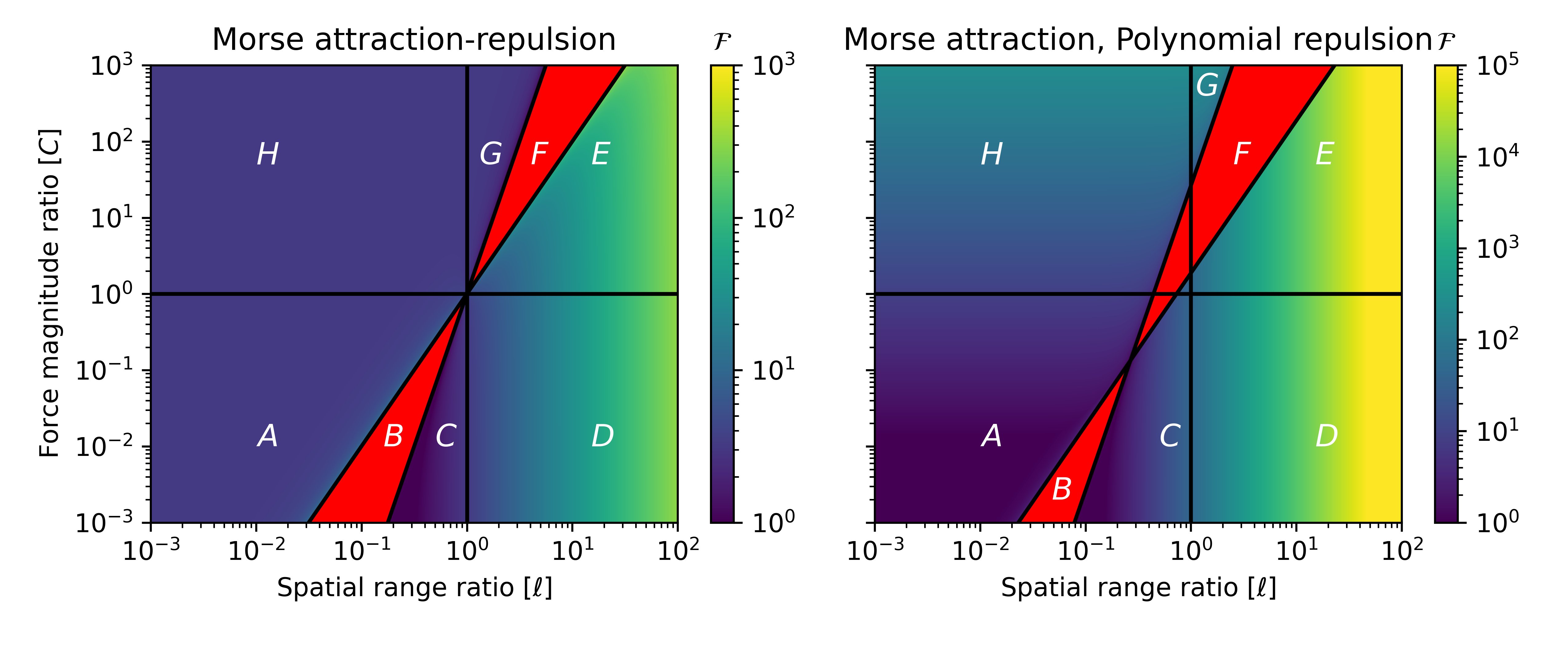}
    \caption{Log-Log plots of the $C\ell$ plane, indicating regions A--H in which distinct behaviors are obtained. $C=R/A$ is the ratio of force magnitudes and $\ell=a/r$ is the ratio of spatial ranges of the attraction and repulsion.
    The heatmap indicates the dimensionless cluster size, $\mathcal{F}$, defined in Eqn.~\eqref{Eqn:ClusterRadiusF}. Red regions correspond to absence of real solutions for $\mathcal{F}$, meaning that no cell cluster of fixed size can exist.
    {\bf Left:} The $C\ell$ plane for the Morse potential as defined in equation~\eqref{eq:MorsePotential1}.
    The boundaries of regions are the curves $C=1, \ell=1, C=\ell^2, C=\ell^4$.
    {\bf Right:} The $C\ell$ plane for
    a Morse attraction and polynomial core-repulsive potential.
     In Region~E, $\mathcal{F}$ increases as the ratio $\ell$ increases. In Region~E near its boundary with region~F,  $\mathcal{F}$ can grow beyond the color scale since Eqn.~\eqref{Eqn:ClusterRadiusF} is near singular.
        }\label{Fig:ClusterSizes}
\end{figure}
Fig.~\ref{Fig:ClusterSizes} shows the $C\ell$ plane, with the dimensionless cluster size $\mathcal{F}$ indicated as a heat map.

In some biological situations, the repulsive length scale $r$ represents a single cell's radius, for example, when cells have an exclusion volume.
In that case, a meaningful multicellular cluster requires  $\mathcal{F}>1$, whereas $\mathcal{F}< 1$ is not meaningful biologically. In this interpretation, we observe that most regions,
outside Region~E, have very small clusters. Fig.~\ref{Fig:ClusterSizes} suggests that the largest clusters are observed for parameters in Region~E near the border $C=\ell^2$ or for larger values of $\ell$, i.e.\ the longer ranged the attraction force the large the cell cluster.


\subsubsection{Morse forces and existence of cluster}
\begin{figure}
     \centering
    \includegraphics[width=0.8\textwidth]{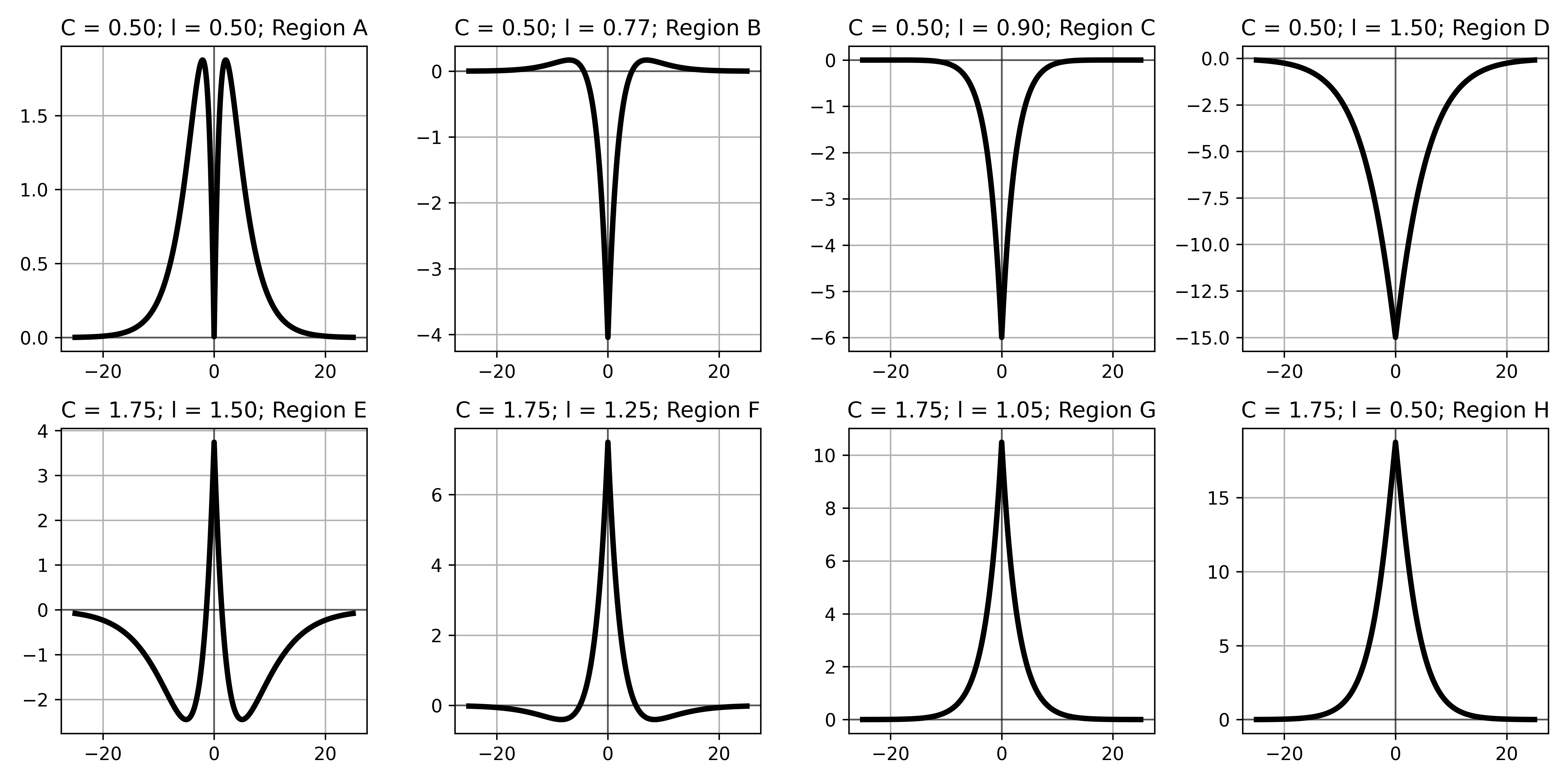}
    \caption{The shapes of the attraction-repulsion Morse potential kernels $K(x)$
    (defined in equation~\eqref{eq:MorsePotential1})
    for $C, \ell$ parameters corresponding to regions A-H in the $C\ell$ plane shown in Fig.~\ref{Fig:ClusterSizes}.
    The kernel parameters in each panel are: Region~A: $A = 5, R = 2.5, a = 1.5, r = 3$; Region~B: $A = 5, R= 2.5, a = 2.3, r = 3$; Region~C: $A = 5, R = 2.5, a = 2.7, r = 3$; Region~D: $A = 5, R = 2.5, a = 4.5, r = 3.0$; Region~E: $A = 5, R = 8.75, a = 4.5, r = 3$; Region~F: $A = 5, R = 8.75, a = 3.75, r = 3.0$;
    Region~G: $A = 5, R = 8.75, a = 3.15, r = 3$; Region~H: $A = 5, R = 8.75, a = 1.5, r = 3.0$.
    }\label{Fig:SimulationKernels}
\end{figure}

Having identified eight regions, bounded by curves shown in Fig.~\ref{Fig:ClusterSizes}, we examine the potentials
induced by a given cell (at ``$x=0$'') on neighboring cells. We display the shapes of these nonlocal 
potential kernels in Fig.~\ref{Fig:SimulationKernels}. The forces ($F=-K_x$) are gradients of the potentials shown in Fig.~\ref{Fig:SimulationKernels}.
(Our later simulations will correspond to the parameter settings shown in this figure).

We summarize the forces and the steady state cluster results in the various parameter regimes as follows:

\begin{description}

    \item[Regions A and C] No cluster possible; $C, \ell < 1$, which implies $R < A$ (attractive force stronger than repulsion) and  $a < r$ (repulsion range greater than attraction range). There is pure repulsion.

    \item[Regions B and F] In region B, $1>\ell^2>C>\ell^4, \ell<1$ implies that $R<A, r<a$, so there is pure attraction. In region F, $\ell^4>C>\ell^2>1, \ell>1$ implies that $R>A, r>a$ and there is or pure repulsion. We showed that in both these regions, no real solutions exist to the steady state cluster problem.

    \item[Region~D] $C < 1$ and $\ell > 1$ which means that
    $R < A$ (attraction is stronger) and $a > r$ (attraction range is longer
    than repulsion range). Density ``blows up'' and no finite-density cluster is possible.

    \item[Region E] $C>1$,  $C<\ell^2, \ell>1$ (so $R>A, a>r$). There is strong local repulsion and long-ranged attraction. Real solutions exist, so cluster formation is possible. However, stability of the cluster is to be determined.

      \item[Region G] $C>\ell^4, \ell>1$. Real solutions exist, cluster formation is possible. However, stability of the cluster is to be determined.

    \item[Region~H] $\ell < 1$, so $a<r$. No cluster can form since repulsion range exceeds attraction range. Cells will
    be repelled from one another.

\end{description}

The above analysis suggests that parameter regions G and E are consistent with cluster formation. Intuition suggests that region~G
is less likely to support stable clusters, since attraction is weak there. Indeed, linear stability analysis will demonstrate that the homogeneous steady state is stable in this region. Our subsequent numerical
simulations (Fig.~\ref{Fig:KymographsSims} and Fig.~\ref{Fig:ProfileSims}) will
confirm these ideas.

\subsubsection{Stability of the HSS, local problem}
From the general conditions \eqref{eq:LSAcondsOn_a's}, we find similar sets of conditions for the Morse potential. When the HSS is unstable, a finite cluster can form.
For the local approximation, we obtain the same inequalities on $a_0, a_2$.

\subsubsection{Stability of the HSS, nonlocal problem}
For the fully nonlocal problem, the HSS stability is determined by \eqref{eq:NonlocalHSSStabil}.  For the Morse potential \eqref{eq:MorsePotential1},
the Fourier transform that appears in the expression \eqref{eq:NonlocalHSSStabil} is
\beq
\label{eq:Khat1}
\hat{K}(q)= \frac{Rr^2}{r^2q^2+1}-\frac{Aa^2}{a^2q^2+1},
\eeq
which we also rewrite as
\beq
\label{eq:Khat2}
 \hat{K}(q) = \frac{r^2 a^2 q^2 (R - A) + (Rr^2 - Aa^2)}{(r^2q^2 + 1)(a^2 q^2 + 1)}.
\eeq
Now consider the perturbation growth rate $\lambda(q)$ given by \eqref{eq:NonlocalHSSStabil}.
That is,
\begin{equation}
\label{eq:lambda(q)}
 \lambda(q)= \rho_0 q^2 \lb \frac{Aa^2}{a^2q^2+1}- \frac{Rr^2}{r^2q^2+1}   \rb.
\end{equation}
To get patterns growing from a perturbed HSS (instability of the uniform cell density solution), we require that there be some positive wavenumber $q^2>0$ that grows, i.e., such that
\[
\lambda(q)
> 0
\quad\iff\quad
\hat{K}(q) < 0.
\]
There are four cases to consider based on regions in the $C\ell$ plane as in Figs.~\ref{Fig:ClusterSizes} and \ref{Fig:KymographsSims}.
\begin{enumerate}
    \item $C > 1$ and $C>\ell^2$ (Regions F,~G and H).
    Here $\hat{K}(q) > 0$ for all $q$.
    The two conditions imply that the numerator of \eqref{eq:Khat2} is always positive.

    \item  $C > 1$ and $C<\ell^2$ (Region E).
    Then $\hat{K}(q)$ has a negative
    minimum at $q = 0$. Since $\ell > C^{1/2} > C^{1/4}$ we have that:
    \[
        \hat{K}(0) = Rr^2 - Aa^2 < 0,\quad
        \frac{\partial^2\hat{K}}{\partial q^2}\Big|_{q = 0}
            = 2 (Aa^4 - Rr^4) > 0.
    \]

    \item If $C < 1$ and $C<\ell^4$ (Regions~C and D).
    Here $\hat{K}(q)$ has a negative
    minimum at $q = 0$. Since $\ell > C^{1/4} > C^{1/2}$ the same
    conditions as in the previous case hold.

    \item If $C < 1$ and $C>\ell^4$ (Regions~A and B);
    then $\hat{K}(q)$ has a
    minimum at $q_c > 0$.
    We look for a second root to the derivative of $\hat{K}(q)$ which
    is equivalent to solving the quartic.
    \[
        q^4 a^4 r^4 (A - R) + 2q^2 a^2 r^2 (Aa^2 - Rr^2) + Aa^4 - Rr^4 = 0.
    \]
    Let $y = q^2$. Look for a positive discriminant,
    \[
        \Delta = 4 a^4 r^4 AR (r^2 - a^2)^2 > 0,
    \]
    meaning we are guaranteed real solutions. Next apply Franciscus Vieta's rule relating products of roots of a polynomial to its coefficients,
    \[
        y_1 y_2 = \frac{Aa^4 - Rr^4}{a^4 r^4 (A - R)} < 0,
    \]
    where $C < 1$ implies that the denominator is positive, and $C>\ell^4$
    implies
    that the numerator is negative.
    This implies that this quadratic has a positive and negative real root.
    We are interested only in the positive root, which is
    \[
        q_c = \frac{1}{ar} \frac{\lb r^2 \sqrt{R} - a^2 \sqrt{A}\rb^{1/2}}{\lb \sqrt{A} - \sqrt{R}\rb^{1/2}}.
    \]

\end{enumerate}


\begin{figure}[h]
    \centering
    \includegraphics[width=0.8\textwidth]{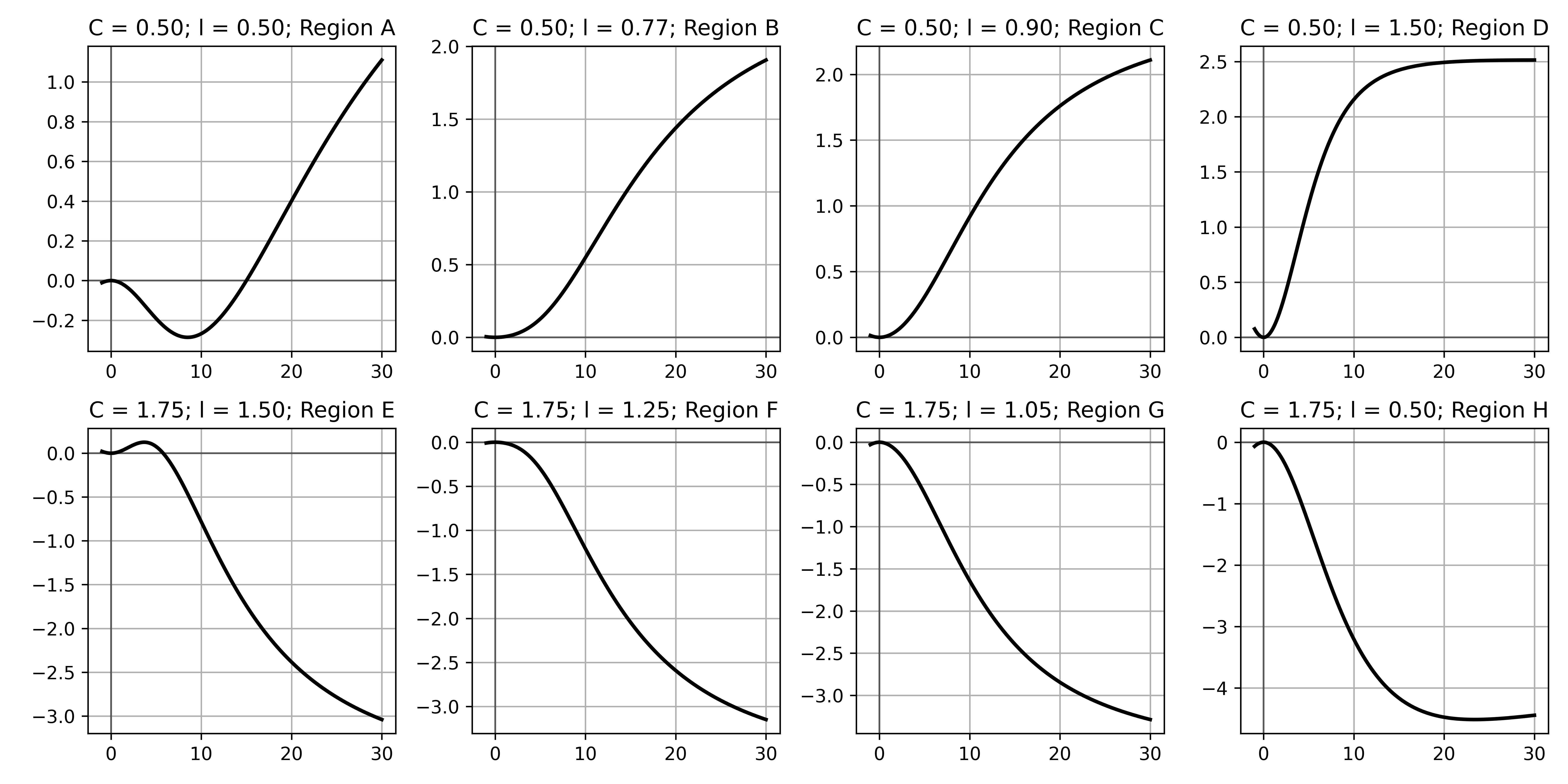}
    \caption{The dispersion relation, showing  $\lambda(q)$ as a function of $q$ from Eqn.~\eqref{eq:lambda(q)}, in the eight regions (A-H) of the $C\ell$ plane shown in Fig.~\ref{Fig:ClusterSizes}. The same parameter values are used as in panels of Fig.~\ref{Fig:SimulationKernels}.   
    }\label{Fig:BifEq}
\end{figure}

The outcome of the linear stability analysis in each parameter regime is now possible. The dispersion relation (a plot of the perturbation growth rate
$\lambda(q)$) is shown in Fig.~\ref{Fig:BifEq}.

\begin{description}

    \item[Regions A and B] Here $\hat{K}(q)$ has a minimum at $q_c \not = 0$, and $\lambda(q)>0$
    for large $q$.

    \item[Regions C and D] $\hat{K}(q)$ has a minimum at $q = 0$ and $\lambda(q) > 0$.

    \item[Region E] $\hat{K}(q)$ has a negative minimum
    at $q = 0$, which results in $\lambda(q) > 0$ in a neighbourhood of zero.
    Thus the homogeneous steady state is unstable, and we expect cluster formation.

    \item[Regions F, G and H] $\hat{K}(q) > 0$ for all $q$ and $\lambda(q) < 0$. The homogeneous steady state
    is stable, and we expect no cluster formation.

\end{description}

Note that the perturbation growth rate, $\lambda(q)$ has positive values for many modes
in regions~A, B, C, and D. This is consistent with results from the
previous section, in which we expect possible blow-up.

To summarize, in Region E of the $C\ell$ plane, the HSS is unstable, and a robust cluster exists. The relevant inequalities that guarantee this cluster are $
C>1,\quad \ell>1, \quad C<\ell^2.$
Note that when the three inequalities $C>1, \ell>1, C<\ell^2$ are satisfied, then the inequality $C<\ell^4$ is automatically satisfied. Hence, in terms of the original Morse parameters,
\begin{equation}
\label{eq:MorseParIneq}
\frac{R}{A}>1,\quad \frac{a}{r}>1, \quad\frac{R}{A}<\frac{a^2}{r^2} \quad \Rightarrow \quad \text{stable cluster exists}.
\end{equation}

\section{Derivation of Morse kernels from chemical signaling}

The Morse potential was introduced in 1926 by Morse to replace the harmonic potential in quantum mechanics systems for diatomic molecules. The Morse potential proved to be more accurate given experimental observations.  Interestingly no derivation of this potential from first principles was provided, to our knowledge, until  \cite{mogilner1995PhDThesis}, and much later in \cite{costa2013morse}.
%
Below, we briefly demonstrate the chemical attraction-repulsion system of \cite{mogilner1995PhDThesis} that directly leads to a Morse type potential.

Following \cite{mogilner1995PhDThesis}, we consider a direct generalization of the model of \cite{keller1971model} that includes chemotaxis up the gradient of a diffusible attractant $c_a(x,t)$, and down the gradient of a diffusible repellent, $c_r(x,t)$, both secreted by the cells. The PDEs describing the cell density $\rho(x,t)$ and the chemical concentrations are

\begin{equation}\label{eq:KS_2PDEs}
\left\{
\begin{split}
    \frac{\partial \rho}{\partial t} &= D_\rho \frac{\partial^2 \rho}{\partial x^2} -  \chi_a\frac{\partial}{\partial x} \lb
    \rho \frac{\partial c_a}{\partial x} \rb
    +  \chi_r\frac{\partial}{\partial x}\lb
   \rho \frac{\partial  c_r}{\partial x} \rb ,\\
   \epsilon_a \frac{\partial c_a}{\partial t}    &= D_a\frac{\partial^2 c_a}{\partial x^2} + s_a\rho - k_a c_a,
    \\
   \epsilon_r \frac{\partial c_r}{\partial t}    &= D_r\frac{\partial^2 c_r}{\partial x^2} + s_r\rho - k_r c_r.
\end{split}
\right.
\end{equation}
Here $D_\rho, D_a, D_r$ are, respectively the cell random motility and the chemical rates of diffusion. The parameters $\chi_a, \chi_r$ are chemotactic coefficients for cells moving towards (away from) the attractant (repellent).  The parameters $s_a, s_r$ are rates of secretion of the two chemicals by cells and $k_a, k_r$ are rates of chemical decay.
The parameters $\epsilon_{a,r}$ are assumed to be small, so that the chemical dynamics operates on a faster timescale than the cell motion.

Signaling molecules diffuse at a far faster rate than the motion of cells. Hence, a quasi steady state (QSS) approximation for the two chemical PDEs can be justified by time-scale separation, implying that

\begin{equation}
\label{eq:QSSchem}
\frac{d^2 c_j}{d x^2}  - \frac{k_j}{ D_j} c_j =-\frac{s_j}{ D_j}\rho, \qquad j=a,r.
\end{equation}
Eqn~\eqref{eq:QSSchem} can be solved by the Green's function method \citep{mogilner1995PhDThesis}, leading to
\begin{equation}
\label{eq:GreenFn}
 c_j(x)= \sqrt{\frac{D_j}{k_j}}\frac{s_j}{2D_j} \int_{-\infty}^{\infty} \exp\lb - \sqrt{\frac{k_j}{D_j}}|x-x'|  \rb \rho(x') dx'   \qquad j=a,r.
\end{equation}
We recognize the above as a convolution of the cell density with a kernel $K_j(x)$ of the form
\begin{equation}
\label{eq:GreenFn2}
K_j(x)= P_j \cdot  p_j\exp\lb - \left|\frac{x}{p_j} \right| \rb, \quad p_j= \sqrt{\frac{D_j}{k_j}}, \quad P_j=\frac{s_j}{2D_j}, \qquad j=a,r.
\end{equation}
Here the expression $p_j$ is the spatial range of the kernel $K_j(x)$ (corresponding to $j=a,r$ for the attraction and repulsion). In our case, there is a superposition of two such terms, one each for the attractant and the repellent. The quantity $P_j$ will prove to be proportional to the amplitudes of the kernel ($A,R$), but we include the chemotactic sensitivities of the cells in those amplitudes, see below.

We use these QSS solutions to eliminate $c_a, c_r$ from the PDE system, but first we rewrite the cell PDE in the compressed form
\begin{equation} \label{eq:Shadowsys01}
   \frac{\partial \rho}{\partial t} = D_\rho \frac{\partial^2 \rho}{\partial x^2} - \frac{\partial}{\partial x} \lb
    \rho \lb - \frac{\partial W}{\partial x} \rb \rb, \quad W=[\chi_r c_r-\chi_a c_a].
\end{equation}
We chose signs above so that $W$ can be interpreted as a chemical potential; for example, if there is no repellent ($c_r=0$) then there would be a ``potential well'' where the attractant is most concentrated.

The form of the expressions in \eqref{eq:GreenFn2} results in spatial ranges of the repellent and attractant, and attraction repulsion force magnitudes
\begin{equation}
\label{eq:MorseChemParams}
r=\sqrt{\frac{D_r}{k_r}}, \quad
a= \sqrt{\frac{D_a}{k_a}}, \quad
R= \frac{\chi_r s_r}{2D_r}, \quad A= \frac{\chi_a s_a}{2D_a}.
\end{equation}
Then the chemical potential can be written as
\begin{equation}
\label{eq:DefnOfPotential}
W(x)=K*\rho, \quad\text{where}\quad K(x)=R r\exp\lb - \left|\frac{x}{r}\right|\rb - A a \exp\lb - \left|\frac{x}{a}\right| \rb.
\end{equation}
The above implies that $W$ acts as a net ``chemical potential landscape''  induced by the cell density due to chemical signaling. This potential sets up the attraction-repulsion force-field that results in cell motion. (As the cells move, that potential landscape will shift.)

Substituting the values of $r,a,R,A$ from \eqref{eq:MorseChemParams} into the conditions \eqref{eq:MorseParIneq} for a stable cluster, we find that
\[
\frac{\chi_r s_r D_a}{\chi_a s_a D_r} >1,\quad
\sqrt{\frac{D_a k_r}{D_r k_a}}>1,
\quad
\frac{\chi_r s_r }{\chi_a s_a } <\frac{ k_r}{ k_a}.
\]
These can be combined into a single statement, namely that for a robust cluster to form,
\begin{equation}
\label{eq:NiceInequal}
\frac{D_r}{D_a}<
\frac{\chi_r s_r}{\chi_a s_a} <\frac{ k_r}{ k_a}.
\end{equation}
The inequality \eqref{eq:NiceInequal} is the only requirement for the existence of a (stable) cluster.

If we normalize all parameters by the rates for the attractant, then we find that the relative diffusion of the repellent should be smaller than the product of cell's repulsion taxis rate and its repellent secretion rate. The repellent relative decay rate should be larger than both of the above. In other words, for a cell cluster to form, the repellent should diffuse slowly and decay rapidly so that its spatial range is small; the cell should have a repulsion taxis and repellent secretion rate that are neither too large nor too small relative to the attractant.


We can see from the above that the magnitudes of repulsion/attraction forces $R,A$ can be ``controlled'' by rates of secretion $s_j$ of the repellent and attractant as well as the cell's chemotactic sensitivity to gradients of each chemical, i.e. the parameters $\chi_j$. The spatial ranges of the interactions, $r,a$ are set by the diffusivity and the decay rates of the chemicals. The lower the rate of diffusion, the smaller the spatial range, and the larger the magnitude of the respective attraction-repulsion force.
The decay rate can be adjusted if cells also secrete enzymes that degrade the ligands. For example, the social amoebae Dictyostelium discoideum secrete an enzyme, phosphodiesterase, that degrades the  attractant cAMP.



\section{Generalization to non-Morse repulsive potentials}

While the above theory proves particularly ``elegant'' for Morse potentials, it has a straightforward generalization to other cases, where the repulsion is not due to chemical signaling. We discuss one example, in which the repulsion is due to a power-law force term, analogous to the concave down portion of a double-well potential. Suppose the repulsive potential is given by
\begin{equation}
\label{eq:NewRepulPot}
K_\text{repul}(x)= r R \phi(x/r),\quad \mbox{where } \phi(x/r)=
\left(\left(\frac{x}{r}\right)^2-1\right)^2, \quad |x|<r
\end{equation}
 where $r$ is the spatial range and $R$ the force magnitude, as before. This is a localized compact repulsion, akin to a non-rigid volume exclusion.

We can find the analogous PDE approximation as before, by  computing the moments of the repulsive potential \eqref{eq:NewRepulPot}, to be
\[
\hat{a_0}=\int_{-r}^r K(x) dx = \gamma_0 R r^2, \quad \hat{a_2}=\int_{-r}^r x^2 K(x) dx=\gamma_2 R r^4,
\]
where some constants, ($\gamma_0=16/15, \gamma_2=8/105$) now appear from the polynomial integration step, as shown in the Appendix~\ref{sec:Moments}.
We observe that the results are similar to those of Morse kernels, save for some multiplicative constants.

Now combining this repulsion with the previously described chemoattractant attractive potential, we obtain similar features, slightly modified by the presence of two constants. For example, we find that the spatial frequencies, previously given by \eqref{eq:Frequencies} are now given by
\begin{equation}
\label{eq:NewFreq}
\omega = \sqrt{\frac{a_0}{a_2}}= \sqrt{\frac{\gamma_0R r^2 - 2A a^2}{\gamma_2R r^4 - 2A a^4}},
\quad
\tilde{\omega}=\sqrt{\frac{\gamma_0 C - 2\ell^2}{\gamma_2 C - 2\ell^4}}.
\end{equation}
(The factors of 2, present in $a_0, a_2$ for the double-Morse potential case in \eqref{eq: MorseLocalCoeffs} no longer cancel out, and hence appear only in the attraction.)
The dimensionless cluster size is $\mathcal{F}=\pi/\tilde{\omega} $.
The analysis carries over, but the boundaries of the behavioural regions in the $C\ell$ plane are now defined by curves $C=2\ell^2/\gamma_0$, $C=2\ell^4/\gamma_2$, etc.
We can see the effect of this distinct repulsive potential by considering the $C\ell$ plane on a log-log plot where the curves are of the form
\[
\log(C)= 2 \log(\ell)- G_0, \quad \log(C)= 4 \log(\ell)- G_2
\]
where $G_0=\log(\gamma_0/2), G_2=\log(\gamma_2/2)$ are merely constants that shift the boundary curves up or down.
%
(See right panel of Fig.~\ref{Fig:ClusterSizes}.)
%
%
%
The new repulsive potential also results in changing of the size of the cluster (where a cluster exists), shown by the somewhat distinct hues on that panel compared to the double-Morse potential case.

More generally, one can consider a repulsive potential of the form
\[
K_\text{general}= r R \, \phi (x/r) 
\]
for some generic function $\phi$.
Provided the function $\phi$ is even, integrable on $-\infty<x<\infty$, the same general conclusions hold, (given that $\phi(x)$ is normalized) but with somewhat specific constants $\gamma_0, \gamma_2$ that result in shifting boundary curves up or down in the $C\ell$ plane. In the above example, the repulsive potential has compact support $-r<x<r$, but this need not be the case in general, as we have seen with Morse repulsion.


\section{Numerical simulation results}

So far, the local and nonlocal Morse model analysis informs us about conditions for instability of a uniform density solution (HSS) and about conditions for existence of a finite cluster. However, we still lack two important pieces of information: (1) Conditions that guarantee a \textbf{stable} cluster and (2) an assessment of the accuracy of the local approximation of the full nonlocal problem.

We sought to address these issues by numerically solving both the local and the nonlocal models in 1D, with the Morse potential specified in \Eref{eq:MorsePotential1}. For the non-local model, we used a domain far larger than the expected cluster size and periodic boundary conditions.
We simulate the problem
%
\beq\label{eq:Carrillo1clusterIPDE1D2}
    \frac{\partial \rho}{\partial t}= 
    \frac{\partial}{\partial x} \left( \rho \frac{\partial (K * \rho)}{\partial x}\right),
\eeq
with $K$ the Morse potential whose kernel is given by
\eqref{eq:MorsePotential1}.
For details on the numerical methods see Appendix~\ref{app:numerics}.

\subsection{Nonlocal model simulations}

\begin{figure}[h]\centering
    \includegraphics[width=\textwidth]{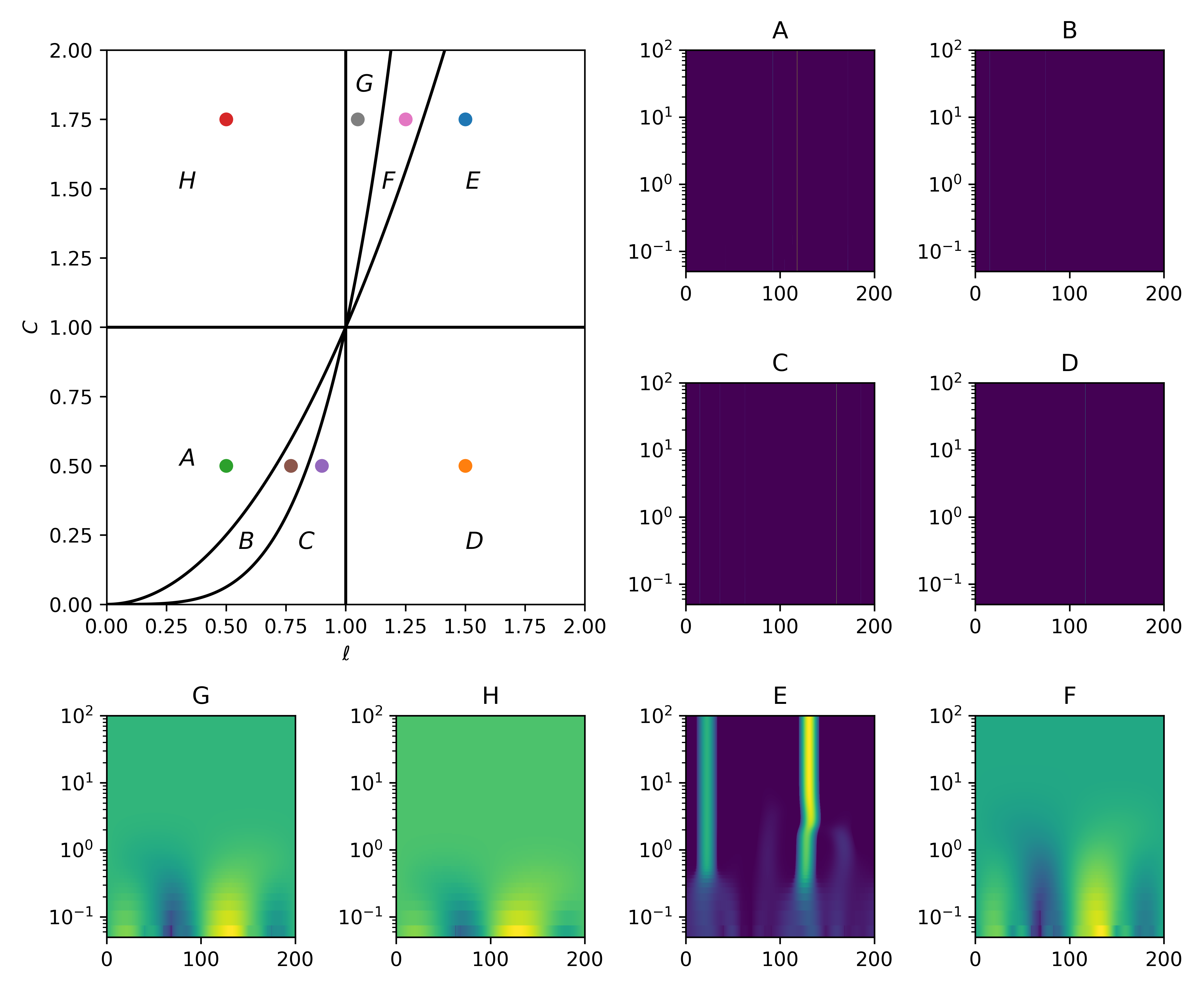}
    \caption{Top left: the $C\ell$ parameter plane showing regions bounded by the curves $C=1, \ell=1, C=\ell^2, C=\ell^4$. These form boundaries of eight distinct regimes of behaviour. (A-F) Simulation results for the 1D nonlocal model~\eqref{eq:Carrillo1clusterIPDE1D2} for an attraction-repulsion Morse Potential kernel with parameters $C,\ell$ shown by corresponding labels in the $C\ell$ plane. (The parameter values are identical to those of Fig.~\ref{Fig:SimulationKernels} and \ref{Fig:BifEq}.) Kymographs show time (vertically upwards on a log plot) and space (horizontal axis).
    The initial condition parameter is $\rho_0 = 20$ in each simulation panel.
    For the final solution profiles at $t=100$, see Fig.~\ref{Fig:ProfileSims}.
    }\label{Fig:KymographsSims}
\end{figure}

We sampled parameter settings corresponding to the eight distinct regions in the $C\ell$ parameter plane shown (top left) in Fig.~\ref{Fig:KymographsSims}, and ran simulations of the nonlocal model ~\eqref{eq:Carrillo1clusterIPDE1D2} in each region, with values (labeled A-H) corresponding to the dots with similar labels in the $C\ell$ plane. All simulations are carried out on a periodic domain. For more 
details we refer to appendix~\ref{app:numerics}.

\begin{figure}[h]\centering
    \includegraphics[width=\textwidth]{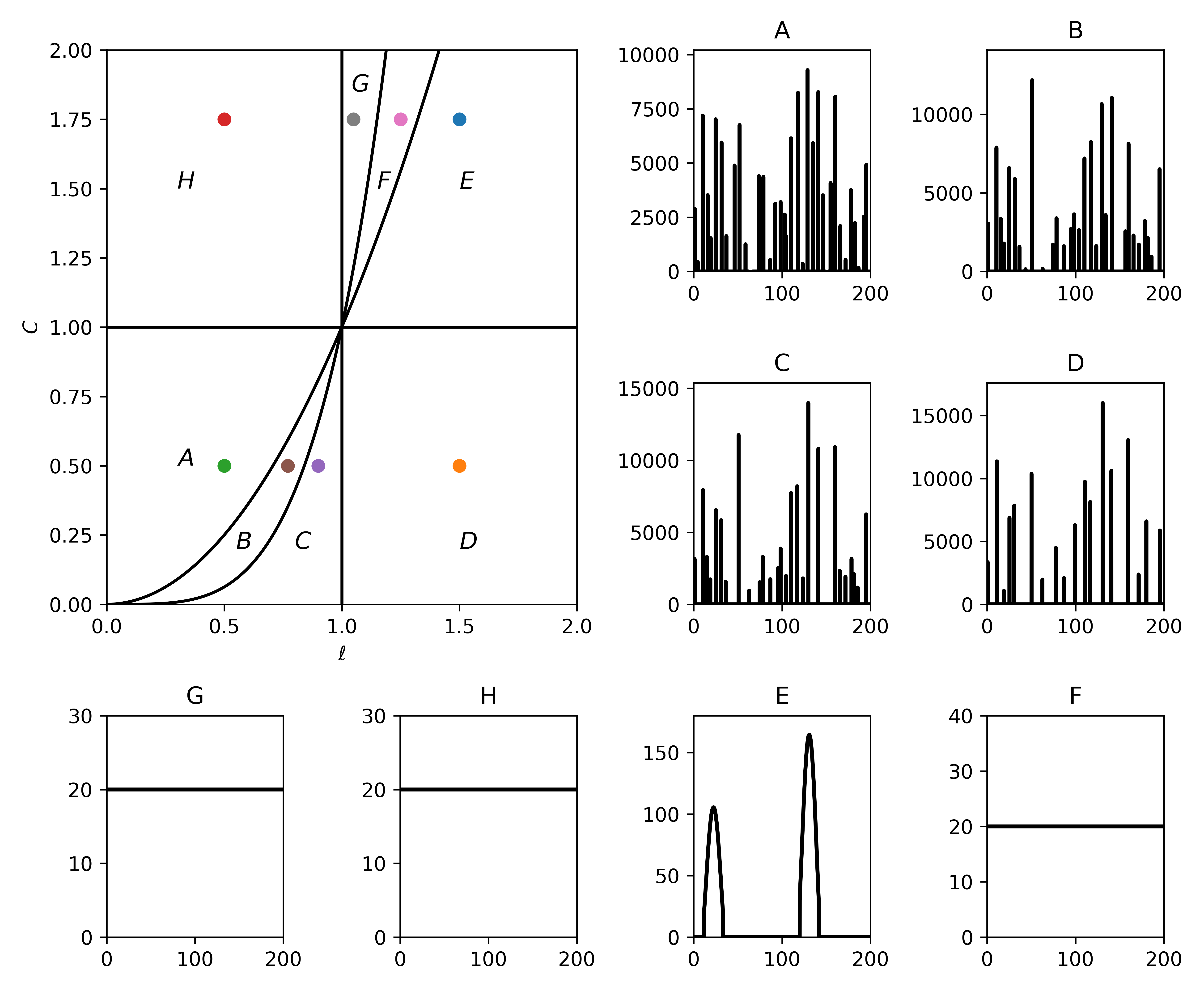}
    \caption{Steady state cluster profiles corresponding to the simulations whose dynamics are shown as kymographs in Fig.~\ref{Fig:KymographsSims}. 
    }\label{Fig:ProfileSims}
\end{figure}

Results are shown as kymographs with time on the vertical log scale in Fig.~\ref{Fig:KymographsSims}. The corresponding non-local kernels for the simulations are as shown in
Fig.~\ref{Fig:SimulationKernels}. The final solution profiles corresponding to the results in Fig.~\ref{Fig:KymographsSims} are then shown in Fig.~\ref{Fig:ProfileSims}. 

\subsubsection{Classification of behaviour}

To summarize the results, we find the following predictions,
\begin{description}

    \item[Regions~A, B and C:] $C < 1, \ell < 1$ ($R < A$ and  $r > a$). As seen from Fig.~\ref{Fig:SimulationKernels}, attraction dominates near each cell (there is a potential well close to the origin) and  repulsion dominates farther away. As we transition from region A to B to C, the repulsion loses strength. In Fig.~\ref{Fig:KymographsSims} the clusters are not very apparent but  the final solution profiles in Fig.~\ref{Fig:ProfileSims} display some small foci of adherent cells that do not aggregate. This region has little biological relevance, and is not consistent with formation of robust clustering.

    \item[Region~D:] $C < 1,\ell >1$ ($R < A$ and $r < a$). Attraction is always stronger than repulsion as shown in Fig.~\ref{Fig:SimulationKernels}, so cells will keep getting closer and closer, implying blow-up to $\delta$-like peaks in the continuum model. This is shown in Fig.~\ref{Fig:ProfileSims}. These peaks are so localized that they do not show up in the heatmaps of Fig.~\ref{Fig:KymographsSims}. 

    \item[Regions~G, F and E:]  $C > 1,\ell > 1$. ($R > A$ and $r < a$). There is short-ranged repulsion and long-ranged attraction. Cells will aggregate and form cluster(s) with finite density. However, in regions G and F, those clusters are unstable, and the final profiles in Fig.~\ref{Fig:ProfileSims} are uniform. In region E, robust clusters can form as shown in Fig~\ref{Fig:KymographsSims} and Fig.~\ref{Fig:ProfileSims}.
 Note that as we transition from region E to F to G, the attraction loses strength, as observed in     Fig.~\ref{Fig:SimulationKernels}.

    \item[Region~H:] $C > 1, \ell < 1$ ($R > A$ and $r > a$).  Repulsion extends further and has greater magnitude than attraction at all distances so cells are repelled from one another and no aggregates form.

\end{description}

\subsection{Simulations of the local approximation}

\begin{figure}[h]
\centering
    \includegraphics[width=\textwidth]{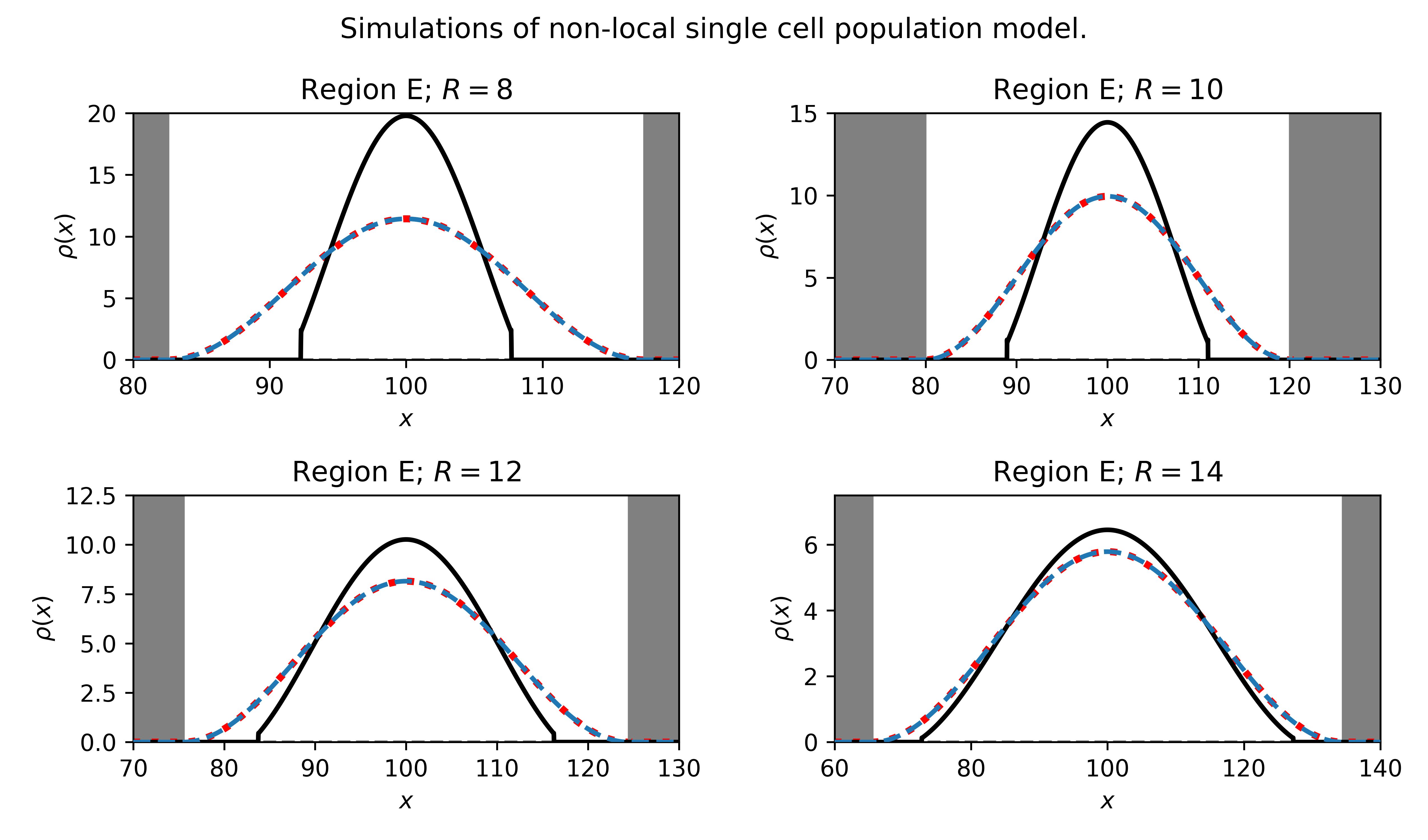}
    \caption{\textbf{Comparison of nonlocal and local models:}
    Solid black: the final numerical solution of the non-local equation~\eqref{eq:Carrillo1clusterIPDE1D2}. Dashed blue: the final numerical solution of the local PDE~\eqref{eq:DimlessAproxPDE1D}. Dotted
    red: the analytically computed steady-state solution~\eqref{eq: rho-C-ell} of the local PDE.
    All parameters are in Region~E of the $C\ell$ plane. We increase the kernel's repulsive strength $R$ (thus increasing $C$), i.e.\ moving in a vertical direction through region~E toward the boundary curve $C = \ell^2$. 
    We note that the local approximation gives an upper bound on the
    cluster radius, and the local approximation improves as the repulsive
    strength increases, when the density of cells in a cluster is lower.
    The white region denotes $[-b, b]$.
    Parameter values: $M = 200$,
    $a =4, A = 1, r = 1$.
    The initial condition for both the local and nonlocal numerical 
    simulation is the local steady-state.
    }\label{fig: RegionComp}
\end{figure}

One of our goals in this paper is to compare the full nonlocal model with its local approximation. Hence, we next simulate the PDE
\beq
\label{eq:DimlessAproxPDE1D}
\frac{\partial \rho}{\partial t}= 
    \frac{\partial}{\partial x} \left( \rho
    \frac{\partial}{\partial x}\lb a_0 \rho + a_2 \rho_{xx} \rb \right),
\eeq
where $a_i$ are rescaled moments of the Morse kernel, that is
\[
       a_0=2(C -\ell^2), \quad            a_2=2(C-\ell^4).
\]

Results are shown in \Fref{fig: RegionComp}. 
We plot the numerically simulated cell density profile of the nonlocal model (black), and compare it to the numerically simulated local model (dashed blue) and the analytically predicted steady-state solution (dotted red). The predicted cluster radius is indicated by the width of the white region.
We find that the numerical and analytical solutions of the local PDE match one another closely, as seen by the overlapping blue and red curves. However, we find that the local PDE approximation is accurate only in some parameter regimes. In the top two panels, we see that the nonlocal model predicts sharp cluster edges and significant compression compared with the local PDE predictions.  
 Hence the analysis of the local approximation does not exactly match the behaviour of the full nonlocal model. The predicted shape becomes a better approximation as $R$ is increased.


\section{Discussion}

In this paper, we have addressed several questions. First, we examined one example of cellular signaling via diffusible chemotactic chemicals that leads to effective nonlocal cell-cell interactions. This example was first proposed by \cite{mogilner1995PhDThesis}, but has not yet been recognized in recent nonlocal modeling efforts. We showed that when cells secrete a diffusible chemical that decays, and that attracts or repels the cells via the process of chemotaxis, it essentially sets up a ``chemical potential'' of the Morse type. 

Morse potentials have been used in more general settings of cell-cell
mechanical interactions in computational biology papers. Morse potentials appear in the context of cell adhesion models \citep{armstrong2006continuum}, as briefly surveyed in Appendix ~\ref{sec:adhes}.
\cite{liu2004coupling}
used a Morse potential to model adhesive/repulsive interactions of red blood
cells (RBCs)  in an advanced computational model of RBC aggregation using an
immersed finite element method. \cite{yazdani2017general} used a modified Morse
potential to represent adhesion of platelets to one another in an ABM model for
blood clotting. The Morse
potential was adapted to experimental data on shear conditions.
\cite{koyama2023effective} used cell tracking in mouse embryos and C.\ elegans
to infer effective mechanical forces and pairwise potential of cell-cell
interactions. They showed that Morse potentials were generally a good fit,
since tuning the Morse parameters allowed for a range of interactions. They
also showed that species differences that implied distinct potential profiles
also accounted for tight vs loose cell aggregate morphologies. As discussed in
\cite{newman2005modeling}, many forms of potentials are reasonable, provided
they account for four parameters for magnitudes and spatial ranges of
attraction and repulsion, akin to our $A, R, a, r$. Hence, even without the
direct link to mechanism, Morse potentials have proven to be convenient
approximations for a variety of non-local interactions.

A second question that we addressed is what kinds of attributes chemotactic signals should have to create a robust cell cluster. 
We showed that the behaviour of cells with attractant and repelent interactions can be classified in a $C\ell$ parameter plane where these dimensionless parameters are ratios of magnitudes and spatial ranges of the attractant and repellent. Similar curves were found in \cite{mogilner1995PhDThesis} in connection with pattern formation criteria. We extended that theory to predict the size of the stable cluster in terms of the chemical properties. We also found the variety of possible behaviours for regions other than that of the cluster formation. In most such regions, the cells cannot aggregate due to dominance of repulsion or the cells collapse due to dominance of attraction.

Some of the theory and predictions here and in \cite{Falco2023} rely on the ability to approximate a nonlocal problem by a local PDE, where steady state solutions can be written explicitly and conveniently (at least in 1D). We quantified the degree of accuracy of such local approximations by numerically simulating and comparing the predictions of the nonlocal problem and its local PDE approximation.

Another question of interest is how the continuum results compare with agent based models. Analytic results for the case of a well-spaced swarm (where all agents are at equal distances from neighbors) in \cite{mogilner2003mutual}
arrived at similar regime boundaries in the $C\ell$ plane, but did not explore the regime of a loose cluster of finite radius.
In \cite{dOrsogna2006self}, the authors investigated an agent-based model for velocity-limited particles that interact with Morse attraction-repulsion kernels. They classified behaviours into ``H-stable'' (well-spaced group) and ``catastrophic'' (collapse into a tight cluster), on a parameter plane very similar to the $C\ell$ plane shown here.
The authors of \cite{leverentz2009asymptotic} consider asymptotic dynamics of a 1D swarming model with attraction-repulsion kernels, including the Morse kernel.
They show how properties of the kernel's moments account for whether the group will contract or spread, or form a steady-state cluster.
In \cite{vecil2013numerical}, the authors include self-propulsion, friction, together with attractive/repulsive Morse
potential.
They assume that the total mass stays constant, regardless of the size of the flock. They numerically simulate their model in 3D, and find regimes for clumps
spheres, dispersion, mills, rigid-body rotation, flocks.
The above examples have features that go beyond our continuum models, and yet some of the properties that we find persist in these distinct situations.


In future plans for the analysis of collective behaviour, and particularly for the migration of the presypathetic ganglion (PSG) of \cite{Kasemeier2015}, it would be important to consider 
the interactions of a second cell type (as modeled in \cite{Falco2023}) with similar signaling mechanism, as well as the migration of a cluster towards its target. In such extensions, some analysis is available, though results are not as easily stated, as many new parameters enter the models.

\begin{appendices}

\section{Numerical methods}\label{app:numerics}

In this section, we give an overview of the numerical methods we employ
to solve both the non-local model and its local approximation. In both cases,
we solve the partial differential equation
\begin{equation}\label{app:model_solve}
    \frac{\partial\rho}{\partial t} =
        -\frac{\partial}{\partial x}\lb \rho v \rb.
\end{equation}
We solve this equation on a large periodic domain $[-L, L]$. We discretize the
equation using a finite volume approach. (See \cite{Gerisch2001} for the theoretical basis.) The non-negative initial condition is 
\[
    \rho(x, 0) = \rho_0 + \sum_{j=1}^{50} \xi_j \sin\lb\frac{2 \pi j x}{L}\rb,
\]
where $\xi_j$ is a uniformly distributed random variable between $(-5, 5)$.
Note that since equation~\eqref{app:model_solve} is hyperbolic (i.e.\ 
does not contain a smoothing diffusion term) the initial condition must be 
smooth.

In detail, the spatial term is discretized
using a third order upwind scheme with a van Leer flux limiter. It remains
to compute the velocities. We consider two cases for the velocities.

\begin{enumerate}
    \item In the non-local case we have
    \[
        v_{\text{non-local}} = - \frac{\partial}{\partial x} (K * \rho).
    \]
    The non-local kernel is discretized using the approach outlined in \cite{Gerisch2010nonlocal}. Most crucially this allows us to use the
    FFT to evaluate the convolution.

    \item In the local case we have
    \[
        v_{\text{local}} = -\frac{\partial}{\partial x}\lb a_0 \rho + a_2 \rho_{xx} \rb.
    \]
    The first term is discretized as usual, the challenging term here is the
    second order derivative term. We discretize this term using a second order
    centered finite difference scheme. This discretization will effectively lead
    to a second order discretization of a fourth order derivative. For our numerical
    calculations it is important not to choose the spatial discretization size
    too small due to the high amplification of round-off errors of this fourth
    order derivative approximation. We estimate the optimal spatial discretization
    size as follows
    \[
        \text{Approximation Error} \sim C_0 h^2 + \frac{C_1}{h^4}
    \]
    where the first term is the local truncation error and the second term is
    the rounding error. Minimizing this expression we obtain an optimal
    discretization size of $10^{-3}$. Most importantly, discretization 
    sizes below this optimal value lead to the emergence of large round-off errors and are to be avoided.
\end{enumerate}

This method of lines scheme results in a large system of nonlinear ordinary
differential equations which we integrate using ROWMAP \citep{Weiner1997a}.
Finally we remark that we extensively tested our numerical method, and in
particular it passes the test cases presented in \cite{bessemoulin2012finite}.

\section{Moments of polynomial repulsion}
\label{sec:Moments}

The moments of the above potential are needed for the local PDE approximation,  namely
\[
\hat{a_0}= r R  \int_{-r}^r \phi(x)
\, dx
= 2r R  \int_{0}^r \left(\frac{x^4}{r^4} -2 \frac{x^2}{r^2}+1\right) \, dx = 2r R \left(\frac{r^5}{5r^4} -2 \frac{r^3}{3r^2}+r\right)
\]
Simplifying leads to $\hat{a_0}=\gamma_0 r^2 R$ where $\gamma_0 = (16/15)$. Similarly the second moment is
\[
\hat{a_2}= r R  \int_{-r}^r x^2 \phi(x)
\, dx= 2r R  \int_{0}^r \left(\frac{x^6}{r^4} -2 \frac{x^4}{r^2}+x^2\right) \, dx =
2r R  \left(\frac{r^7}{7r^4} -2 \frac{r^5}{5r^2}+\frac{r^3}{3}\right).
\]
Simplifying results in $\hat{a_2}=\gamma_2 r^4 R$ with $\gamma_2=(8/105)$.
Thus, the moments of the repulsion are similar to those of the Morse repulsive potential, but with some new constants $\gamma_0,\gamma_2$ appearing.

\section{Morse kernels in adhesion models}
\label{sec:adhes}

Exponential kernels appear in other classes of  theoretical models, including non-local adhesion models. For instance, the model proposed
by \cite{armstrong2006continuum} for cell density $u(x,t)$ is
\beq\label{eq:AdhesionModel1}
    u_t = Du_{xx} - \alpha \lb \frac{u}{R} \int_{-R}^{R} u(x + r) \Omega(r) \dd r \rb_x,
\eeq
where $R$ is the cell's sensing radius i.e.\ roughly the distance around the cell
over which cell-cell forces are exerted. See also \cite{painter2010impact,painter2015nonlocal,buttenschon2018space}.
One common choice of the integral kernel is the exponential
\[
    \Omega = \frac{r}{\abs{r}} \omega_0 \exp\lb -\frac{r}{\xi} \rb,
\]
where $\omega_0$ is a normalization constant. Following the steps in \cite{buttenschon2020non}, we
can write \eqref{eq:AdhesionModel1} in potential form i.e.\
\beq\label{eq:AdhesionModel2}
    u_t = Du_{xx} - \alpha \nabla \lb \frac{u}{R} \nabla \int_{-R}^{R} u(x + r) V(r) \dd r \rb,
\eeq
where
\[
    V(r) = \xi \exp\lb-\frac{r}{\xi}\rb.
\]

In other words, non-local adhesion models (or dispersal models) using Laplace
kernels are a type of Morse potential. 

\end{appendices}


\bibliography{refs} 


\begin{thebibliography}{39}
\ifx \bisbn   \undefined \def \bisbn  #1{ISBN #1}\fi
\ifx \binits  \undefined \def \binits#1{#1}\fi
\ifx \bauthor  \undefined \def \bauthor#1{#1}\fi
\ifx \batitle  \undefined \def \batitle#1{#1}\fi
\ifx \bjtitle  \undefined \def \bjtitle#1{#1}\fi
\ifx \bvolume  \undefined \def \bvolume#1{\textbf{#1}}\fi
\ifx \byear  \undefined \def \byear#1{#1}\fi
\ifx \bissue  \undefined \def \bissue#1{#1}\fi
\ifx \bfpage  \undefined \def \bfpage#1{#1}\fi
\ifx \blpage  \undefined \def \blpage #1{#1}\fi
\ifx \burl  \undefined \def \burl#1{\textsf{#1}}\fi
\ifx \doiurl  \undefined \def \doiurl#1{\url{https://doi.org/#1}}\fi
\ifx \betal  \undefined \def \betal{\textit{et al.}}\fi
\ifx \binstitute  \undefined \def \binstitute#1{#1}\fi
\ifx \binstitutionaled  \undefined \def \binstitutionaled#1{#1}\fi
\ifx \bctitle  \undefined \def \bctitle#1{#1}\fi
\ifx \beditor  \undefined \def \beditor#1{#1}\fi
\ifx \bpublisher  \undefined \def \bpublisher#1{#1}\fi
\ifx \bbtitle  \undefined \def \bbtitle#1{#1}\fi
\ifx \bedition  \undefined \def \bedition#1{#1}\fi
\ifx \bseriesno  \undefined \def \bseriesno#1{#1}\fi
\ifx \blocation  \undefined \def \blocation#1{#1}\fi
\ifx \bsertitle  \undefined \def \bsertitle#1{#1}\fi
\ifx \bsnm \undefined \def \bsnm#1{#1}\fi
\ifx \bsuffix \undefined \def \bsuffix#1{#1}\fi
\ifx \bparticle \undefined \def \bparticle#1{#1}\fi
\ifx \barticle \undefined \def \barticle#1{#1}\fi
\bibcommenthead
\ifx \bconfdate \undefined \def \bconfdate #1{#1}\fi
\ifx \botherref \undefined \def \botherref #1{#1}\fi
\ifx \url \undefined \def \url#1{\textsf{#1}}\fi
\ifx \bchapter \undefined \def \bchapter#1{#1}\fi
\ifx \bbook \undefined \def \bbook#1{#1}\fi
\ifx \bcomment \undefined \def \bcomment#1{#1}\fi
\ifx \oauthor \undefined \def \oauthor#1{#1}\fi
\ifx \citeauthoryear \undefined \def \citeauthoryear#1{#1}\fi
\ifx \endbibitem  \undefined \def \endbibitem {}\fi
\ifx \bconflocation  \undefined \def \bconflocation#1{#1}\fi
\ifx \arxivurl  \undefined \def \arxivurl#1{\textsf{#1}}\fi
\csname PreBibitemsHook\endcsname

\bibitem[\protect\citeauthoryear{Armstrong
  et~al.}{2006}]{armstrong2006continuum}
\begin{barticle}
\bauthor{\bsnm{Armstrong}, \binits{N.J.}},
\bauthor{\bsnm{Painter}, \binits{K.J.}},
\bauthor{\bsnm{Sherratt}, \binits{J.A.}}:
\batitle{A continuum approach to modelling cell--cell adhesion}.
\bjtitle{Journal of theoretical biology}
\bvolume{243}(\bissue{1}),
\bfpage{98}--\blpage{113}
(\byear{2006})
\end{barticle}
\endbibitem

\bibitem[\protect\citeauthoryear{Bessemoulin-Chatard and
  Filbet}{2012}]{bessemoulin2012finite}
\begin{barticle}
\bauthor{\bsnm{Bessemoulin-Chatard}, \binits{M.}},
\bauthor{\bsnm{Filbet}, \binits{F.}}:
\batitle{A finite volume scheme for nonlinear degenerate parabolic equations}.
\bjtitle{{SIAM} Journal on Scientific Computing}
\bvolume{34}(\bissue{5}),
\bfpage{559}--\blpage{583}
(\byear{2012})
\end{barticle}
\endbibitem

\bibitem[\protect\citeauthoryear{Byrne and Drasdo}{2009}]{byrne2009individual}
\begin{barticle}
\bauthor{\bsnm{Byrne}, \binits{H.}},
\bauthor{\bsnm{Drasdo}, \binits{D.}}:
\batitle{Individual-based and continuum models of growing cell populations: a
  comparison}.
\bjtitle{Journal of mathematical biology}
\bvolume{58},
\bfpage{657}--\blpage{687}
(\byear{2009})
\end{barticle}
\endbibitem

\bibitem[\protect\citeauthoryear{Buttensch{\"o}n and
  Edelstein-Keshet}{2020}]{buttenschon2020bridging}
\begin{barticle}
\bauthor{\bsnm{Buttensch{\"o}n}, \binits{A.}},
\bauthor{\bsnm{Edelstein-Keshet}, \binits{L.}}:
\batitle{Bridging from single to collective cell migration: A review of models
  and links to experiments}.
\bjtitle{PLoS computational biology}
\bvolume{16}(\bissue{12}),
\bfpage{1008411}
(\byear{2020})
\end{barticle}
\endbibitem

\bibitem[\protect\citeauthoryear{Buttensch{\"o}n and
  Hillen}{2021}]{buttenschon2020non}
\begin{botherref}
\oauthor{\bsnm{Buttensch{\"o}n}, \binits{A.}},
\oauthor{\bsnm{Hillen}, \binits{T.}}:
Non-local cell adhesion models: Steady states and bifurcations
(2021)
\end{botherref}
\endbibitem

\bibitem[\protect\citeauthoryear{Buttensch{\"o}n
  et~al.}{2018}]{buttenschon2018space}
\begin{barticle}
\bauthor{\bsnm{Buttensch{\"o}n}, \binits{A.}},
\bauthor{\bsnm{Hillen}, \binits{T.}},
\bauthor{\bsnm{Gerisch}, \binits{A.}},
\bauthor{\bsnm{Painter}, \binits{K.J.}}:
\batitle{A space-jump derivation for non-local models of cell--cell adhesion
  and non-local chemotaxis}.
\bjtitle{Journal of mathematical biology}
\bvolume{76},
\bfpage{429}--\blpage{456}
(\byear{2018})
\end{barticle}
\endbibitem

\bibitem[\protect\citeauthoryear{Bonner}{1998}]{bonner1998way}
\begin{barticle}
\bauthor{\bsnm{Bonner}, \binits{J.T.}}:
\batitle{A way of following individual cells in the migrating slugs of
  dictyostelium discoideum}.
\bjtitle{Proceedings of the National Academy of Sciences}
\bvolume{95}(\bissue{16}),
\bfpage{9355}--\blpage{9359}
(\byear{1998})
\end{barticle}
\endbibitem

\bibitem[\protect\citeauthoryear{Bernoff and
  Topaz}{2016}]{bernoff2016biological}
\begin{barticle}
\bauthor{\bsnm{Bernoff}, \binits{A.J.}},
\bauthor{\bsnm{Topaz}, \binits{C.M.}}:
\batitle{Biological aggregation driven by social and environmental factors: A
  nonlocal model and its degenerate cahn--hilliard approximation}.
\bjtitle{{SIAM} Journal on Applied Dynamical Systems}
\bvolume{15}(\bissue{3}),
\bfpage{1528}--\blpage{1562}
(\byear{2016})
\end{barticle}
\endbibitem

\bibitem[\protect\citeauthoryear{Chitnis et~al.}{2012}]{chitnis2012building}
\begin{barticle}
\bauthor{\bsnm{Chitnis}, \binits{A.B.}},
\bauthor{\bsnm{Dalle~Nogare}, \binits{D.}},
\bauthor{\bsnm{Matsuda}, \binits{M.}}:
\batitle{Building the posterior lateral line system in zebrafish}.
\bjtitle{Developmental neurobiology}
\bvolume{72}(\bissue{3}),
\bfpage{234}--\blpage{255}
(\byear{2012})
\end{barticle}
\endbibitem

\bibitem[\protect\citeauthoryear{Costa~Filho et~al.}{2013}]{costa2013morse}
\begin{barticle}
\bauthor{\bsnm{Costa~Filho}, \binits{R.N.}},
\bauthor{\bsnm{Alencar}, \binits{G.}},
\bauthor{\bsnm{Skagerstam}, \binits{B.-S.}},
\bauthor{\bsnm{Andrade}, \binits{J.S.}}:
\batitle{Morse potential derived from first principles}.
\bjtitle{Europhysics Letters}
\bvolume{101}(\bissue{1}),
\bfpage{10009}
(\byear{2013})
\end{barticle}
\endbibitem

\bibitem[\protect\citeauthoryear{Chaplain et~al.}{2020}]{chaplain2020bridging}
\begin{barticle}
\bauthor{\bsnm{Chaplain}, \binits{M.A.}},
\bauthor{\bsnm{Lorenzi}, \binits{T.}},
\bauthor{\bsnm{Macfarlane}, \binits{F.R.}}:
\batitle{Bridging the gap between individual-based and continuum models of
  growing cell populations}.
\bjtitle{Journal of Mathematical Biology}
\bvolume{80}(\bissue{1}),
\bfpage{343}--\blpage{371}
(\byear{2020})
\end{barticle}
\endbibitem

\bibitem[\protect\citeauthoryear{D’Orsogna et~al.}{2006}]{dOrsogna2006self}
\begin{barticle}
\bauthor{\bsnm{D’Orsogna}, \binits{M.R.}},
\bauthor{\bsnm{Chuang}, \binits{Y.-L.}},
\bauthor{\bsnm{Bertozzi}, \binits{A.L.}},
\bauthor{\bsnm{Chayes}, \binits{L.S.}}:
\batitle{Self-propelled particles with soft-core interactions: patterns,
  stability, and collapse}.
\bjtitle{Physical review letters}
\bvolume{96}(\bissue{10}),
\bfpage{104302}
(\byear{2006})
\end{barticle}
\endbibitem

\bibitem[\protect\citeauthoryear{Elliott and Garcke}{1996}]{elliott1996cahn}
\begin{barticle}
\bauthor{\bsnm{Elliott}, \binits{C.M.}},
\bauthor{\bsnm{Garcke}, \binits{H.}}:
\batitle{On the cahn--hilliard equation with degenerate mobility}.
\bjtitle{{SIAM} journal on mathematical analysis}
\bvolume{27}(\bissue{2}),
\bfpage{404}--\blpage{423}
(\byear{1996})
\end{barticle}
\endbibitem

\bibitem[\protect\citeauthoryear{Falc{\'o} et~al.}{2023}]{Falco2023}
\begin{botherref}
\oauthor{\bsnm{Falc{\'o}}, \binits{C.}},
\oauthor{\bsnm{Baker}, \binits{R.E.}},
\oauthor{\bsnm{Carrillo}, \binits{J.A.}}:
A local continuum model of cell-cell adhesion.
{SIAM} Journal on Applied Mathematics,
17--42
(2023)
\end{botherref}
\endbibitem

\bibitem[\protect\citeauthoryear{Friedl and Mayor}{2017}]{friedl2017tuning}
\begin{barticle}
\bauthor{\bsnm{Friedl}, \binits{P.}},
\bauthor{\bsnm{Mayor}, \binits{R.}}:
\batitle{Tuning collective cell migration by cell--cell junction regulation}.
\bjtitle{Cold Spring Harbor perspectives in biology}
\bvolume{9}(\bissue{4}),
\bfpage{029199}
(\byear{2017})
\end{barticle}
\endbibitem

\bibitem[\protect\citeauthoryear{Friedl et~al.}{1995}]{friedl1995migration}
\begin{barticle}
\bauthor{\bsnm{Friedl}, \binits{P.}},
\bauthor{\bsnm{Noble}, \binits{P.B.}},
\bauthor{\bsnm{Walton}, \binits{P.A.}},
\bauthor{\bsnm{Laird}, \binits{D.W.}},
\bauthor{\bsnm{Chauvin}, \binits{P.J.}},
\bauthor{\bsnm{Tabah}, \binits{R.J.}},
\bauthor{\bsnm{Black}, \binits{M.}},
\bauthor{\bsnm{Z{\"a}nker}, \binits{K.S.}}:
\batitle{Migration of coordinated cell clusters in mesenchymal and epithelial
  cancer explants in vitro}.
\bjtitle{Cancer research}
\bvolume{55}(\bissue{20}),
\bfpage{4557}--\blpage{4560}
(\byear{1995})
\end{barticle}
\endbibitem

\bibitem[\protect\citeauthoryear{Gerisch and
  Chaplain}{2008}]{gerisch2008mathematical}
\begin{barticle}
\bauthor{\bsnm{Gerisch}, \binits{A.}},
\bauthor{\bsnm{Chaplain}, \binits{M.A.}}:
\batitle{Mathematical modelling of cancer cell invasion of tissue: local and
  non-local models and the effect of adhesion}.
\bjtitle{Journal of theoretical biology}
\bvolume{250}(\bissue{4}),
\bfpage{684}--\blpage{704}
(\byear{2008})
\end{barticle}
\endbibitem

\bibitem[\protect\citeauthoryear{Gerisch}{2001}]{Gerisch2001}
\begin{botherref}
\oauthor{\bsnm{Gerisch}, \binits{A.}}:
Numerical methods for the simulation of taxis diffusion reaction systems.
PhD thesis,
Martin-Luther-Universitat Halle-Wittenberg
(2001)
\end{botherref}
\endbibitem

\bibitem[\protect\citeauthoryear{Gerisch}{2010}]{Gerisch2010nonlocal}
\begin{barticle}
\bauthor{\bsnm{Gerisch}, \binits{A.}}:
\batitle{On the approximation and efficient evaluation of integral terms in pde
  models of cell adhesion}.
\bjtitle{IMA Journal of Numerical Analysis}
\bvolume{30}(\bissue{1}),
\bfpage{173}--\blpage{194}
(\byear{2010})
\end{barticle}
\endbibitem

\bibitem[\protect\citeauthoryear{Kasemeier-Kulesa et~al.}{2015}]{Kasemeier2015}
\begin{barticle}
\bauthor{\bsnm{Kasemeier-Kulesa}, \binits{J.C.}},
\bauthor{\bsnm{Morrison}, \binits{J.A.}},
\bauthor{\bsnm{Lefcort}, \binits{F.}},
\bauthor{\bsnm{Kulesa}, \binits{P.M.}}:
\batitle{Trkb/bdnf signalling patterns the sympathetic nervous system}.
\bjtitle{Nature communications}
\bvolume{6},
\bfpage{8281}
(\byear{2015})
\end{barticle}
\endbibitem

\bibitem[\protect\citeauthoryear{Koyama et~al.}{2023}]{koyama2023effective}
\begin{barticle}
\bauthor{\bsnm{Koyama}, \binits{H.}},
\bauthor{\bsnm{Okumura}, \binits{H.}},
\bauthor{\bsnm{Ito}, \binits{A.M.}},
\bauthor{\bsnm{Nakamura}, \binits{K.}},
\bauthor{\bsnm{Otani}, \binits{T.}},
\bauthor{\bsnm{Kato}, \binits{K.}},
\bauthor{\bsnm{Fujimori}, \binits{T.}}:
\batitle{Effective mechanical potential of cell--cell interaction explains
  three-dimensional morphologies during early embryogenesis}.
\bjtitle{PLoS Computational Biology}
\bvolume{19}(\bissue{8}),
\bfpage{1011306}
(\byear{2023})
\end{barticle}
\endbibitem

\bibitem[\protect\citeauthoryear{Keller and Segel}{1971}]{keller1971model}
\begin{barticle}
\bauthor{\bsnm{Keller}, \binits{E.F.}},
\bauthor{\bsnm{Segel}, \binits{L.A.}}:
\batitle{Model for chemotaxis}.
\bjtitle{Journal of theoretical biology}
\bvolume{30}(\bissue{2}),
\bfpage{225}--\blpage{234}
(\byear{1971})
\end{barticle}
\endbibitem

\bibitem[\protect\citeauthoryear{Knutsdottir
  et~al.}{2017}]{knutsdottir2017polarization}
\begin{barticle}
\bauthor{\bsnm{Knutsdottir}, \binits{H.}},
\bauthor{\bsnm{Zmurchok}, \binits{C.}},
\bauthor{\bsnm{Bhaskar}, \binits{D.}},
\bauthor{\bsnm{Palsson}, \binits{E.}},
\bauthor{\bsnm{Dalle~Nogare}, \binits{D.}},
\bauthor{\bsnm{Chitnis}, \binits{A.B.}},
\bauthor{\bsnm{Edelstein-Keshet}, \binits{L.}}:
\batitle{Polarization and migration in the zebrafish posterior lateral line
  system}.
\bjtitle{PLoS computational biology}
\bvolume{13}(\bissue{4}),
\bfpage{1005451}
(\byear{2017})
\end{barticle}
\endbibitem

\bibitem[\protect\citeauthoryear{Leverentz
  et~al.}{2009}]{leverentz2009asymptotic}
\begin{barticle}
\bauthor{\bsnm{Leverentz}, \binits{A.J.}},
\bauthor{\bsnm{Topaz}, \binits{C.M.}},
\bauthor{\bsnm{Bernoff}, \binits{A.J.}}:
\batitle{Asymptotic dynamics of attractive-repulsive swarms}.
\bjtitle{{SIAM} Journal on Applied Dynamical Systems}
\bvolume{8}(\bissue{3}),
\bfpage{880}--\blpage{908}
(\byear{2009})
\end{barticle}
\endbibitem

\bibitem[\protect\citeauthoryear{Liu et~al.}{2004}]{liu2004coupling}
\begin{barticle}
\bauthor{\bsnm{Liu}, \binits{Y.}},
\bauthor{\bsnm{Zhang}, \binits{L.}},
\bauthor{\bsnm{Wang}, \binits{X.}},
\bauthor{\bsnm{Liu}, \binits{W.K.}}:
\batitle{Coupling of navier--stokes equations with protein molecular dynamics
  and its application to hemodynamics}.
\bjtitle{International Journal for Numerical Methods in Fluids}
\bvolume{46}(\bissue{12}),
\bfpage{1237}--\blpage{1252}
(\byear{2004})
\end{barticle}
\endbibitem

\bibitem[\protect\citeauthoryear{Mogilner et~al.}{2003}]{mogilner2003mutual}
\begin{barticle}
\bauthor{\bsnm{Mogilner}, \binits{A.}},
\bauthor{\bsnm{Edelstein-Keshet}, \binits{L.}},
\bauthor{\bsnm{Bent}, \binits{L.}},
\bauthor{\bsnm{Spiros}, \binits{A.}}:
\batitle{Mutual interactions, potentials, and individual distance in a social
  aggregation}.
\bjtitle{Journal of mathematical biology}
\bvolume{47},
\bfpage{353}--\blpage{389}
(\byear{2003})
\end{barticle}
\endbibitem

\bibitem[\protect\citeauthoryear{Merchant et~al.}{2018}]{merchant2018rho}
\begin{barticle}
\bauthor{\bsnm{Merchant}, \binits{B.}},
\bauthor{\bsnm{Edelstein-Keshet}, \binits{L.}},
\bauthor{\bsnm{Feng}, \binits{J.J.}}:
\batitle{A rho-gtpase based model explains spontaneous collective migration of
  neural crest cell clusters}.
\bjtitle{Developmental biology}
\bvolume{444},
\bfpage{262}--\blpage{273}
(\byear{2018})
\end{barticle}
\endbibitem

\bibitem[\protect\citeauthoryear{Mogilner}{1995}]{mogilner1995PhDThesis}
\begin{botherref}
\oauthor{\bsnm{Mogilner}, \binits{A.}}:
Modelling spatio-angular patterns in cell biology.
PhD thesis,
University of British Columbia
(1995)
\end{botherref}
\endbibitem

\bibitem[\protect\citeauthoryear{Mar{\'e}e et~al.}{1999}]{maree1999migration}
\begin{barticle}
\bauthor{\bsnm{Mar{\'e}e}, \binits{A.F.}},
\bauthor{\bsnm{Panfilov}, \binits{A.V.}},
\bauthor{\bsnm{Hogeweg}, \binits{P.}}:
\batitle{Migration and thermotaxis of dictyostelium discoideum slugs, a model
  study}.
\bjtitle{Journal of theoretical biology}
\bvolume{199}(\bissue{3}),
\bfpage{297}--\blpage{309}
(\byear{1999})
\end{barticle}
\endbibitem

\bibitem[\protect\citeauthoryear{Murray}{2003}]{murray2003mathematical}
\begin{bbook}
\bauthor{\bsnm{Murray}, \binits{J.D.}}:
\bbtitle{Mathematical Biology: II: Spatial Models and Biomedical Applications}
vol. \bseriesno{18}.
\bpublisher{Springer},
\blocation{New York}
(\byear{2003})
\end{bbook}
\endbibitem

\bibitem[\protect\citeauthoryear{Newman}{2005}]{newman2005modeling}
\begin{botherref}
\oauthor{\bsnm{Newman}, \binits{T.J.}}:
Modeling multi-cellular systems using sub-cellular elements.
arXiv preprint q-bio/0504028
(2005)
\end{botherref}
\endbibitem

\bibitem[\protect\citeauthoryear{Othmer and Hillen}{2002}]{othmer2002diffusion}
\begin{barticle}
\bauthor{\bsnm{Othmer}, \binits{H.G.}},
\bauthor{\bsnm{Hillen}, \binits{T.}}:
\batitle{The diffusion limit of transport equations ii: Chemotaxis equations}.
\bjtitle{{SIAM} Journal on Applied Mathematics}
\bvolume{62}(\bissue{4}),
\bfpage{1222}--\blpage{1250}
(\byear{2002})
\end{barticle}
\endbibitem

\bibitem[\protect\citeauthoryear{Painter et~al.}{2010}]{painter2010impact}
\begin{barticle}
\bauthor{\bsnm{Painter}, \binits{K.J.}},
\bauthor{\bsnm{Armstrong}, \binits{N.J.}},
\bauthor{\bsnm{Sherratt}, \binits{J.A.}}:
\batitle{The impact of adhesion on cellular invasion processes in cancer and
  development}.
\bjtitle{Journal of theoretical biology}
\bvolume{264}(\bissue{3}),
\bfpage{1057}--\blpage{1067}
(\byear{2010})
\end{barticle}
\endbibitem

\bibitem[\protect\citeauthoryear{Painter et~al.}{2015}]{painter2015nonlocal}
\begin{barticle}
\bauthor{\bsnm{Painter}, \binits{K.J.}},
\bauthor{\bsnm{Bloomfield}, \binits{J.}},
\bauthor{\bsnm{Sherratt}, \binits{J.}},
\bauthor{\bsnm{Gerisch}, \binits{A.}}:
\batitle{A nonlocal model for contact attraction and repulsion in heterogeneous
  cell populations}.
\bjtitle{Bulletin of mathematical biology}
\bvolume{77},
\bfpage{1132}--\blpage{1165}
(\byear{2015})
\end{barticle}
\endbibitem

\bibitem[\protect\citeauthoryear{Palsson and Othmer}{2000}]{palsson2000model}
\begin{barticle}
\bauthor{\bsnm{Palsson}, \binits{E.}},
\bauthor{\bsnm{Othmer}, \binits{H.G.}}:
\batitle{A model for individual and collective cell movement in dictyostelium
  discoideum}.
\bjtitle{Proceedings of the National Academy of Sciences}
\bvolume{97}(\bissue{19}),
\bfpage{10448}--\blpage{10453}
(\byear{2000})
\end{barticle}
\endbibitem

\bibitem[\protect\citeauthoryear{Vecil et~al.}{2013}]{vecil2013numerical}
\begin{barticle}
\bauthor{\bsnm{Vecil}, \binits{F.}},
\bauthor{\bsnm{Lafitte}, \binits{P.}},
\bauthor{\bsnm{Linares}, \binits{J.R.}}:
\batitle{A numerical study of attraction/repulsion collective behavior models:
  3d particle analyses and 1d kinetic simulations}.
\bjtitle{Physica D: Nonlinear Phenomena}
\bvolume{260},
\bfpage{127}--\blpage{144}
(\byear{2013})
\end{barticle}
\endbibitem

\bibitem[\protect\citeauthoryear{Weijer}{2009}]{weijer2009collective}
\begin{barticle}
\bauthor{\bsnm{Weijer}, \binits{C.J.}}:
\batitle{Collective cell migration in development}.
\bjtitle{Journal of cell science}
\bvolume{122}(\bissue{18}),
\bfpage{3215}--\blpage{3223}
(\byear{2009})
\end{barticle}
\endbibitem

\bibitem[\protect\citeauthoryear{Weiner et~al.}{1997}]{Weiner1997a}
\begin{barticle}
\bauthor{\bsnm{Weiner}, \binits{R.}},
\bauthor{\bsnm{Schmitt}, \binits{B.A.}},
\bauthor{\bsnm{Podhaisky}, \binits{H.}}:
\batitle{Rowmap--a row-code with krylov techniques for large stiff odes}.
\bjtitle{Applied Numerical Mathematics}
\bvolume{25},
\bfpage{303}--\blpage{319}
(\byear{1997})
\end{barticle}
\endbibitem

\bibitem[\protect\citeauthoryear{Yazdani et~al.}{2017}]{yazdani2017general}
\begin{barticle}
\bauthor{\bsnm{Yazdani}, \binits{A.}},
\bauthor{\bsnm{Li}, \binits{H.}},
\bauthor{\bsnm{Humphrey}, \binits{J.D.}},
\bauthor{\bsnm{Karniadakis}, \binits{G.E.}}:
\batitle{A general shear-dependent model for thrombus formation}.
\bjtitle{PLoS computational biology}
\bvolume{13}(\bissue{1}),
\bfpage{1005291}
(\byear{2017})
\end{barticle}
\endbibitem

\end{thebibliography}

\end{document}